\documentclass[a4paper,fleqn,usenatbib]{mnras}

\usepackage{newtxtext,newtxmath}

\usepackage[T1]{fontenc}

\usepackage{graphicx}	
\usepackage{amsmath}	
\usepackage{longtable}
\usepackage[dvipsnames]{xcolor}

\newcommand{\numstars}{74} 
\newcommand{\numstarsWithCeph}{79} 
\newcommand{\numstarsQuestionable}{34\ } 
\newcommand{\nickelimgs}{691} 
\newcommand{\nickelnights}{46} 
\newcommand{\pipelinename}{PIPS} 
\newcommand{\newstars}{two} 
\graphicspath{{./}{figures/}}
\newcommand{\revised}[1]{\textcolor{black}{#1}}
\newcommand{\revisedII}[1]{\textcolor{black}{#1}}

\title[RR Lyrae stars in M15]{Periods and classifications of RR Lyrae stars in the globular cluster M15}

\author[Hoffman et al.]{Andrew M. Hoffman,$^{1,2,3}$\thanks{E-mail: andrewmh@berkeley.edu}
Yukei S. Murakami,$^{1,2,3}$\thanks{E-mail: sterling.astro@berkeley.edu}
\newauthor
WeiKang Zheng,$^{1}$
Benjamin E. Stahl,$^{1,2,4}$
and Alexei V. Filippenko$^{1,5}$
\\
$^{1}$Department of Astronomy, University of California, Berkeley, CA 94720-3411, USA\\
$^{2}$Department of Physics, University of California, Berkeley, CA 94720-7300, USA\\
$^{3}$Google Lick Predoctoral Fellow\\
$^{4}$Marc J. Staley Graduate Fellow\\
$^{5}$Miller Senior Fellow, Miller Institute for Basic Research in Science, University of California, Berkeley, CA 94720, USA\\
}

\date{Accepted to MNRAS (December 2020).} 

\pubyear{2020}

\begin{document}
\label{firstpage}
\pagerange{\pageref{firstpage}--\pageref{lastpage}}
\maketitle

\begin{abstract}
    We present measurements of the periods, amplitudes, and types of \numstars{} RR Lyrae stars in the globular cluster M15 derived from Nickel 1\,m telescope observations conducted at Lick Observatory in 2019 and 2020. Of these \revised{RR Lyrae stars}, \newstars{} were previously reported but without a determination of the period. In addition, we identify five Type II Cepheid variable stars for which we report three novel period determinations, and a further \numstarsQuestionable stars with uncertain classifications and periods. We discuss the development and subsequent application to our data of a new Python package, Period-determination and Identification Pipeline Suite (\pipelinename{}), based on a new adaptive free-form fitting technique to detect the periods of variable stars with a clear treatment of uncertainties.
\end{abstract}




\section{Introduction} 
    \label{sec:intro}
    Pulsating variable stars periodically change their brightness over timescales ranging from a few \revised{minutes} to a few months, making them members of a select group of astronomical objects that are dynamic over time intervals that are observable by humans \citep[for a review, see, e.g.,][]{Percy_2007_VS_Review}. Moreover, the observable details of this dynamic behaviour follows patterns that are unique to each type of variable star, a property that makes photometric time-series analysis a particularly useful tool for their study \citep{photometry_2007}.
    
    The astrophysical applications of variable-star observations span many fields, such as testing hydrodynamical models \citep{Smolec2012_blazhko_hydro_model} and investigating the expansion of the Universe \citep[][and references therein]{Scolnic2019_NextGenCosmo}. If the measurements of a variable star's \revised{parameters are} uncertain, this error can propagate into many other ``downstream'' facets of an analysis --- accuracy is, therefore, critical. 
    
    RR Lyrae variable stars are evolved low-mass stars which have moved off the main sequence onto the horizontal branch. They are among the most common types of variable stars in \revised{Galactic globular clusters} \citep{Clement_2001}. Despite a large number of samples and a long history of observations (from \citealt{Fleming_1901_RRLyrae_Discovery} to, e.g., \citealt{OGLE_2020}), the behaviour of RR Lyrae stars is not fully understood.
    For instance, the expected long-term behaviour \revised{of the period} is not agreed upon nor immediately evident in observations. Stars do not remain on the horizontal branch forever \revised{\citep{Jurcsik_2003}}, but the degree, timescale, and type of evolution within the branch are not trivial to theorise or measure. Observations spanning large temporal ranges ($\sim 100$\,yr) suggest that RR Lyrae stars have generally stable periods \citep{Arellano_Ferro_2018}, but some theories predict noticeable changes \citep{Fadeyev2018}. Moreover, the short-term modulation of light curves \citep{Blazhko_1907} seen in some stars remains an open question. Recent studies \citep{Szabo__2014} suggest that the multimode property of such ``Blazhko stars'' resembles known multimode RR Lyrae stars (RRd), and studies connecting them are needed.
    
    When investigating RR Lyrae stars it is valuable for repeated observations to be made as often as possible. Such data yield (among other things) a basis with which to investigate long-term changes and may yield insights into \revised{tricky phenomena such as multimode pulsations and the Blazhko effect}. Messier 15 (NGC 7078, M15) is well suited for such studies.  It is an old and star-dense globular cluster \citep{APOGEE_M15_age_2020} that contains many evolved star types, and in particular, dozens to hundreds of RR Lyrae stars which can be well resolved by telescopes of moderate size. Many photometric observations of RR Lyrae stars in M15 can be found in the literature, including \citet{Bailey1919}, \citet{Wemple1932}, \citet{Mannino1956}, \citet{Makarova1965}, \citet{Hogg73}, \citet{Filippenko1981}, \citet{Bingham1984}, \citet{Silbermann1995}, \citet{Corwin_2008}, \citet{Ferro2006}, and \citet{Siegel2015}.

    In this paper, we present new observations of RR Lyrae stars in M15 and an analysis of their types and pulsation periods. In Section \ref{sec: observation and data reduction} we describe our observations and data-processing methods.
    We also discuss matching the variable stars detected from raw data with known catalogued variable stars. Section \ref{sec: analysis} describes our method of detecting the period of observed variable stars, with an emphasis on dealing with widely spaced data, and also discusses our methods of handling error propagation and star classification. In Section \ref{sec: results} we present the periods, magnitudes, and type classification of \numstars{} RR Lyrae stars recovered from our observations and derived by our analysis. We also highlight several stars for which few or no previous detections exist, discuss a number of additional stars for which we were not able to accurately determine periods, and briefly describe period determinations for a small number of Type II Cepheid variable stars. We give conclusions and directions for future work in Section~\ref{sec: long-term}.

\section{Observations and Data Reduction}
\label{sec: observation and data reduction}
    \subsection{Observations}
    \label{sec: observation}

        \begin{figure}
            \centering
            \includegraphics[width=\linewidth]{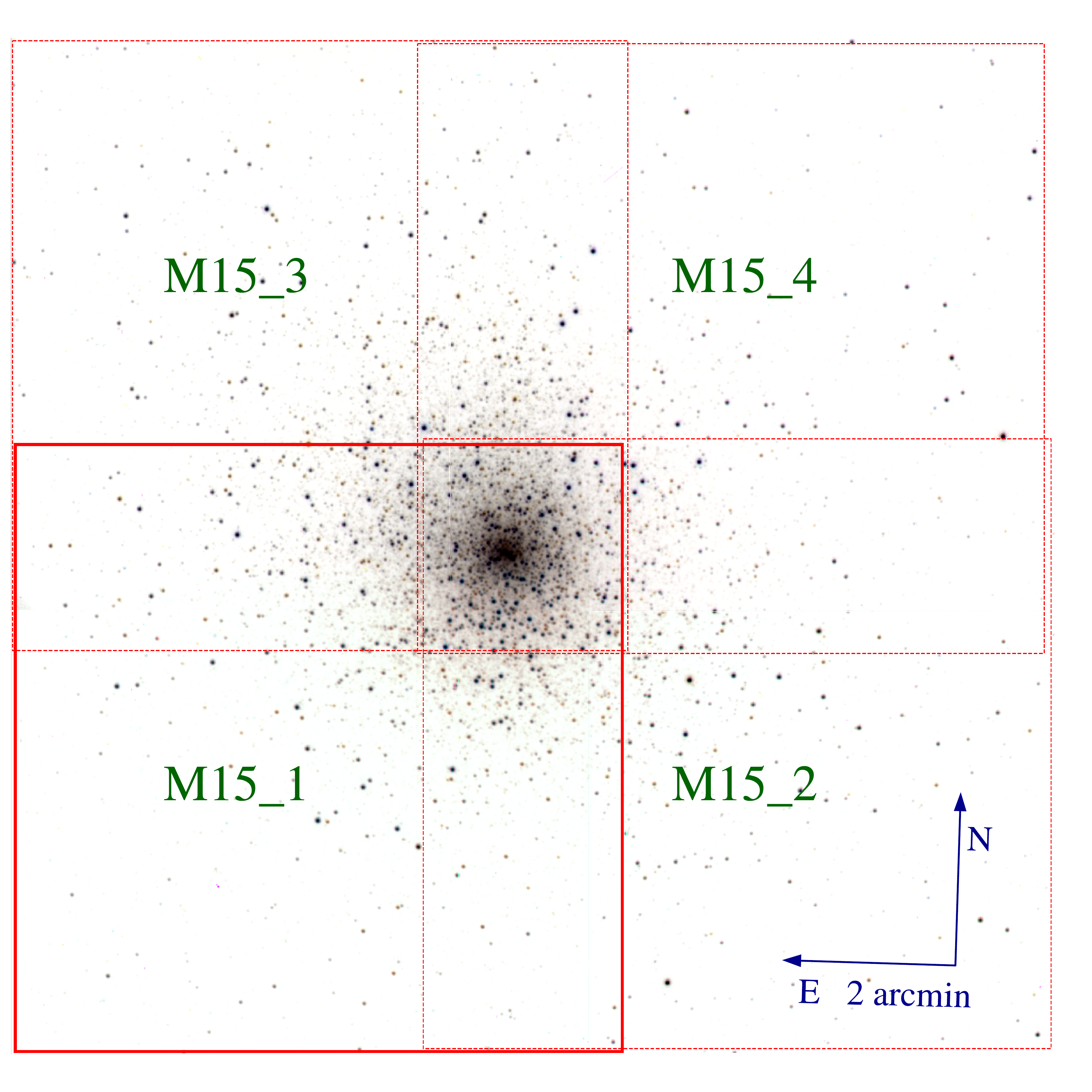}
            \caption{Tiled images of M15, with each being a $BVR$ composite of inverted-color Nickel telescope images. Our observations cover $10'\times10'$ centred at the core of M15 by combining four tiled images (M15\_1 -- M15\_4) that each have a $\sim6'\times6'$ field. Note the slight rotation ($\sim 1.5^\circ$) of the compass rose.}
            \label{fig: M15 fields}
        \end{figure}

        Over approximately one year, we obtained \nickelimgs{} images of M15 on \nickelnights{} nights using the Nickel 1\,m telescope, located at Lick Observatory (Mt. Hamilton, CA). We observed M15 for an average of 20\,min per night, on nights devoted mainly to supernova photometry, with a typical time between observations of roughly 7\,d. As RR Lyrae stars generally have oscillation periods of 0.2--0.9\,d \citep{Corwin_2008}, our observing cadence produces widely scattered data when compared to the pulsation period. Ideally, fast pulsators such as RR Lyrae stars should be observed as often as possible to capture the full range of behaviour \citep{VanderPlas_2018}. With widely scattered data, period determination becomes difficult, as does avoiding aliasing.
        
        After considering the field of view of the Nickel telescope, we chose to take 4 tiled fields of M15 in order to image the largest number of stars. The arrangement is shown in Figure~\ref{fig: M15 fields}. Specific locations of variable stars within these fields can be found in Figures~\ref{fig: M15_1_coord}--\ref{fig: M15_4_coord}. Our typical observations of a given field consisted of exposures of 20, 60, and 180\,s in the $V$ band, generally three times per night. This strategy increases the likelihood of obtaining at least one data point per night with a sufficiently high signal-to-noise ratio (SNR), given the varying brightness of our targets. When time permits, images were also taken in the $B$ band.

        With a plate scale of $0.368''$ pixel$^{-1}$ after binning\footnote{We use $2 \times 2$ binning with the $2048 \times 2048$ CCD whose normal plate scale is $0.184''$ pixel$^{-1}$.}, \revisedII{and a median seeing of $(2.02\pm 0.50)''$,} the core of M15 is sometimes not well resolved in our images. Because of this, there are limitations in period determinations for stars which are located close to the core due to possible contamination and blending. 
        When examining previously published data, we see a similar trend where stars which are harder to individually resolve are less likely to have reported periods \citep[e.g.,][]{Siegel2015}.
        To minimise the effect of this issue, tiled fields are overlapped at the core, increasing the chance of detection by collecting a larger number of data points.
        \revisedII{Although outside the scope of this work, higher resolution images of M15 (such as Hubble photometry e.g., \citealt{Sarajedini2007}) exist, which can help distinguish between stars in these dense regions.}
        
    \subsection{Data Reduction}
        \subsubsection{LOSSPhotPypeline}
             We utilise the LOSSPhotPypeline\footnote{\url{ https://github.com/benstahl92/LOSSPhotPypeline}} \citep[LPP;][]{Stahl_2019} to perform photometry on the images taken from the Nickel telescope and construct light curves in the Nickel2 natural system.
             The LPP provides robust methods for uncertainty calculation, including reduction on simulated stars in the field. We use these uncertainties as a proxy by which to identify poor-quality data that need to be removed from further analysis. As our use of the LPP is consistent with that of \citet{Stahl_2019}, we defer to their Section 3 for a more detailed discussion of its capabilities.

        \subsubsection{Calibration}
            \begin{table}
                \centering
                \caption{Statistics for calibration stars used in each field. $N_\mathrm{stars}$ is the number of calibration stars in each field. $\Delta V$ is the deviation from the corresponding PS1 magnitude and $\sigma V$ is the observed standard deviation, both in the $V$ band.}
                \begin{tabular}{cccccc}
                \hline \hline
                Field &  $N_{\mathrm{stars}}$ & $\overline{\Delta V}$ & $(\Delta V)_\mathrm{max}$ & $\overline{\sigma V}$ & $(\sigma V)_\mathrm{max}$\\
                 & & (mag) & (mag) & (mag) & (mag)\\
                \hline
                M15\_1 &     9 &      $-0.0002$ &      0.0242 &      0.0170 &     0.0319 \\
                M15\_2 &     9 &      $-0.0004$ &      0.0157 &      0.0162 &     0.0303 \\
                M15\_3 &     8 &      $-0.0004$ &      0.0346 &      0.0218 &     0.0340 \\
                M15\_4 &    10 &      $-0.0012$ &      0.0210 &      0.0192 &     0.0245 \\
                \hline \hline
                \end{tabular}
                \label{tab: cal_stats}
            \end{table}
            
            We calibrate each field by picking  bright, nonvariable stars that have minimal (typically $< 0.03$\,mag) deviation from the corresponding PS1 catalogue values, converted to the Landolt-system (\citealt{Landolt1}, \citealt{Landolt2}) using the prescription of \cite{Tonry_etal_Panstars_2012}, and then to the Nickel2 natural system using the transformations and color terms presented by \cite{Stahl_2019}, 
            and small (also typically $< 0.03$\,mag) scatter (per calibration star) in each of their observed magnitudes through the entire time series. The first criterion ensures calibration consistency between the four different fields in our tiling strategy (see Fig.~\ref{fig: M15 fields}), and the second --- in conjunction with the requirement that only those calibration stars that are detected in every image for a given field be used --- ensures a consistent calibration for images taken across long time intervals. \revised{This calibration process introduces scatter in our data. The characteristic value of this scatter for combined fields can be calculated from the mean and maximum standard deviation of calibration star magnitudes $\overline{\sigma V}$. We see a characteristic value of $\langle\overline{\sigma V}\rangle \approx 0.02$ and $\langle(\sigma V)_\mathrm{max}\rangle\approx 0.03$, which is representative of the uncertainty of our measurements. Considering the characteristic uncertainty in our data}, we set a 0.03\,mag floor on magnitude uncertainties (commensurate with our second criterion).
            We summarise important statistics for our calibration choices in Table~\ref{tab: cal_stats}, while detailed information is relegated to Table~\ref{tab: calibration stars} in the Appendix. Unless otherwise noted, we do not accept data above an uncertainty threshold $\sigma_V \ge  \sigma_{V,\mathrm{cut}}$, where $\sigma_{V,\mathrm{cut}} = \overline{\sigma_V} + \mathrm{std}(\sigma_V)/2$ for each star.

        \subsubsection{Cross-matching}
        
            \begin{figure}
                \centering\
                \includegraphics[width=\linewidth]{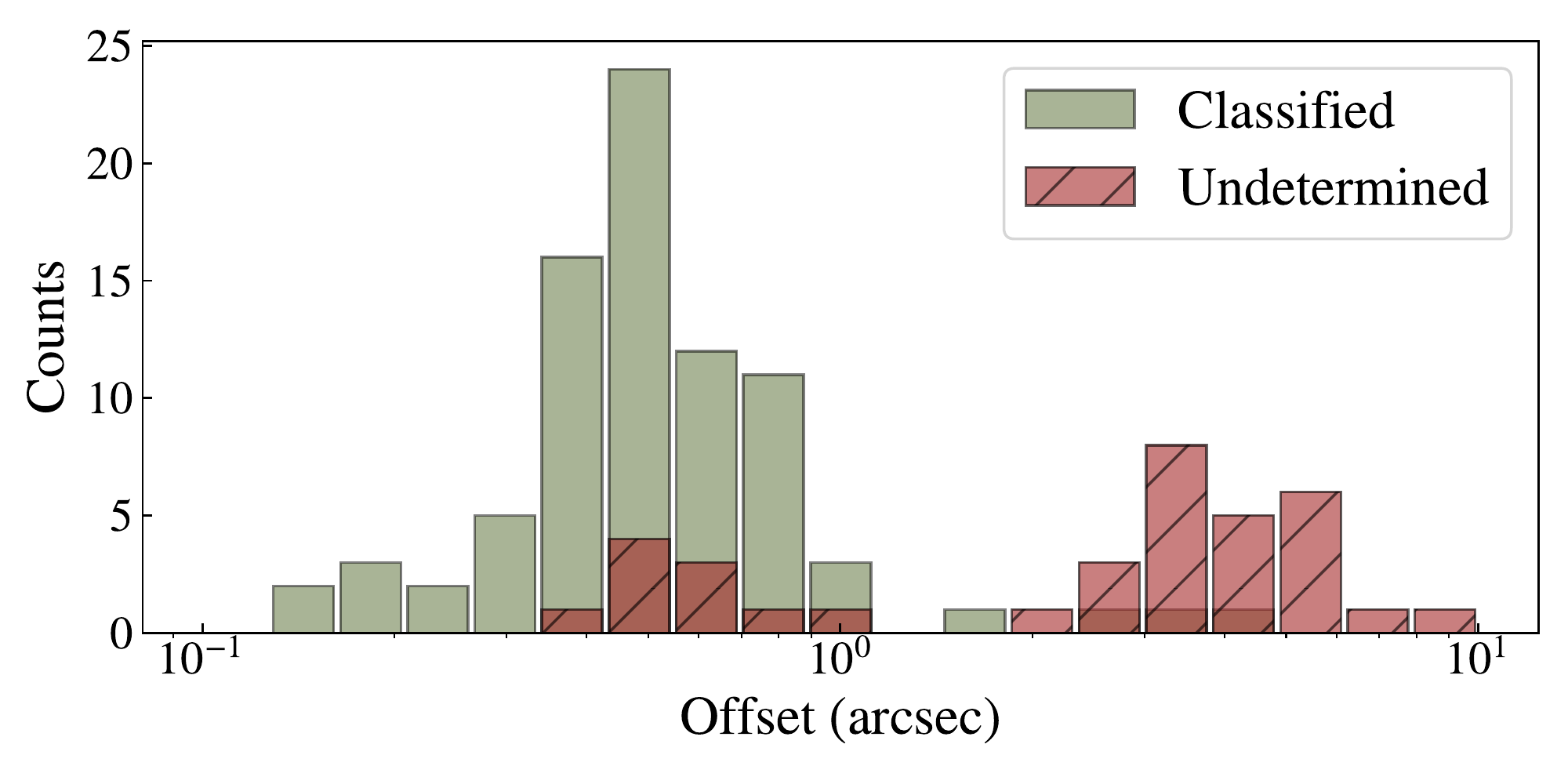}
                \caption{Offset of our candidate stars from GCVS catalog $\theta_i$, in arcsec. Successful classifications tend to correlate with a small offset.}
                \label{fig: distance to GCVS}
            \end{figure}
            
            We identify candidate variable stars using the LPP's ability to automatically detect all resolvable stars in an image. For each field, this procedure identifies $\sim 1300$ candidates, which we then compare to previously published positions for variable stars. Specifically, we calculate the angular distance between star $i$ identified by the LPP and star $j$ in \citet[][hereafter GCVS]{GCVS2017}, 
            $
            \theta_{ij} = \cos^{-1}\big[\sin(\delta_i)\sin(\delta_j)+
             \cos(\delta_i)\cos(\delta_j)\cos(\alpha_i-\alpha_j)\big].
            $
            Here, $\alpha$ is the right ascension and $\delta$ is the declination of the subscripted object, both in radians. Once this two-dimensional array is generated, we take $\min(\theta_j)_i$ to create a list of the variable stars nearest to each candidate. This list contains many duplicates, so we further narrow it by taking $\min(\min(\theta_j)_i)_j$ for each duplicate star $j$. 
            The resulting list contains each star's coordinates ($\alpha$ and $\delta$), identification (ID), and the offset from the relevant GCVS star. This process is repeated for all four fields, M15\_1 to M15\_4. 
            Our star ID is expressed with ``V'' numbers (e.g., V001). This is a slightly modified version of older notations, such as V1, as seen in \citet{Bailey_1902} and other papers. After this process, we obtain $\sim 60$ candidate stars in each field to be processed.
            
            Since there are overlaps between each field, we process stars in more than one field whenever possible, thereby allowing us to choose the data based on various quality parameters (see Sec.~\ref{sec: period detection}). The resulting, cross-matched catalogue shows strong agreement with GCVS coordinates. There are two populations, one with a small ($\sim 0.4''$) offset and one with a larger ($\sim 2.5''$) offset, which are visible in Figure~\ref{fig: distance to GCVS}. Our final selections (see Sec.~\ref{sec: results}) suggest that stars in the second (large-distance) population mostly fail to provide high-quality data. Assuming GCVS has high accuracy, this is the expected result. 
            Note that while most of our successfully classified stars fall in this ``low offset'' population, the offset value $\theta_i$ is used only for cross-matching candidates, and our final classification is based on photometric results (see Sec.~\ref{sec: results}).

\section{Analysis}   
\label{sec: analysis}
    \begin{figure*}
        \centering
        \includegraphics[width=\linewidth]{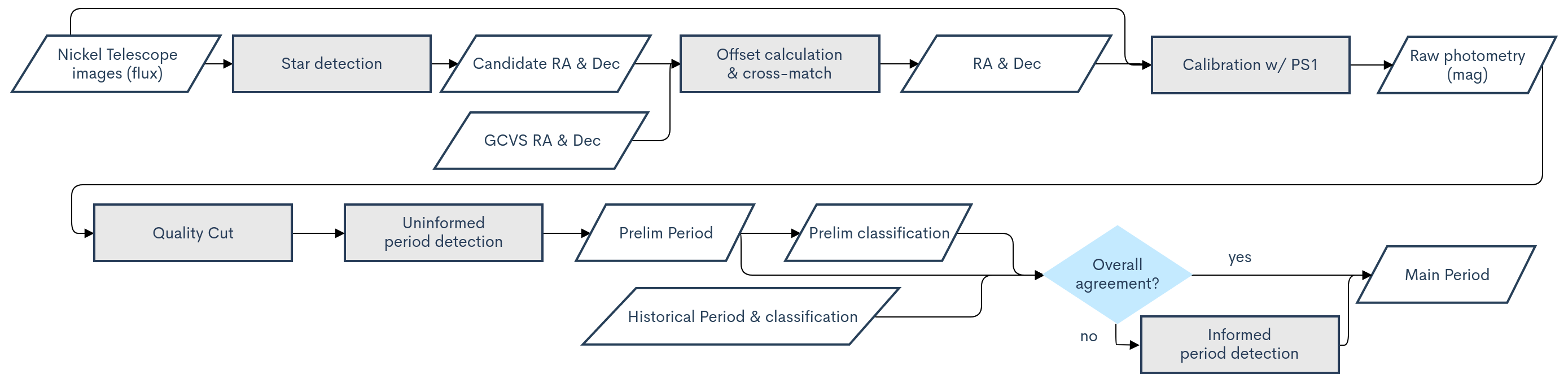}
        \caption{Data-reduction process chart.}
        \label{fig:my_label}
    \end{figure*}
    
    \subsection{\pipelinename{}}
    \label{sec: period detection}

    We have developed the Period-determination and Identification Pipeline Suite (\pipelinename{})\footnote{\url{https://github.com/SterlingYM/PIPS}}, a new Python package to analyse variable-star data. While there are many algorithms and code bases available for this purpose, we found that several improvements were needed to obtain the best results, especially for difficult objects characterised by widely spaced data, low SNR, and few data points.

        \subsubsection{Motivation and Background}
            \paragraph*{Period Uncertainty}
            --- We note that in older publications, particularly in studies which rely on the time of maximum light \citep[see, e.g.,][]{Bailey_1902,Wemple1932,Makarova1965}, uncertainty quantification is not treated carefully. More recently, the method of fitting to folded data instead of relying on the maximum or minimum times has become more prominent \citep[see][]{GaiaDR2_variability}, and with it, a clearer understanding and treatment of uncertainty. This is of great importance for modern studies as long-term data can and will be used to study the evolution of certain quantities (e.g., period and metallicity), and a careful treatment of uncertainty is required for proper error propagation and the statements of confidence that such calculations support. In developing \pipelinename{}, our primary concern was therefore to examine the uncertainty created when making period determinations.
            
            \paragraph*{Template Bias}
            --- When analysing variable-star photometry, our goal is to determine the period and create analytic functions (e.g., Fourier series) which provide an accurate description of the star's light curve. Although high-cadence observations can yield a complete light curve and through it, a direct measurement of the period, data are often\footnote{For instance, in many modern all-sky surveys.} taken with a lower cadence \revised{which does not immediately show a clear morphology}\revisedII{\ --- this is indeed the case with our data}. This necessitates \revised{a phase-folded }light curve \citep{VanderPlas_2018}, and therefore requires a determination of the period before analysis can proceed. The principal challenge in our work is then to \revisedII{make accurate period determinations, and hence, construct accurate phase-folded light curves.}
            
            The Lomb-Scargle periodogram is the most common algorithm designed to detect sinusoidal signals in widely or unevenly spaced data \citep{VanderPlas_2018}. Template Fourier Fitting \citep[TFF;][]{Kovacs_Kupi_2007_TFF} is a technique which leverages high-cadence data of similar phenomena to create an informed prior for light-curve fitting. 
            Both of these techniques can make period determinations, but the bias induced by one's choice of light-curve template is not widely addressed. 
            
            Owing to Lomb-Scargle's nature as a correlation function to a sinusoidal pattern \citep{Lomb1976,Scargle1982}, taking the maximum returned power value only yields the best-fit period to a sinusoidal wave. Because \revised{the higher order terms of a Fourier fitting do not always have negligible amplitude}, this sinusoidal dependence is analogous to the template dependence in TFF. 
            Both the template dependence in TFF and the sinusoidal dependence in LS skew results from these methods.
            The RRab type of RR Lyrae stars, for instance, have sharply peaked maxima which are poorly described by a sinusoid \citep[e.g., light curves in][]{Filippenko1981,Corwin_2008}. 
            For this reason, we employ a template-free fitting method with \pipelinename{} to minimise bias.
            
            \begin{figure}
                \centering
                \includegraphics[width=\linewidth]{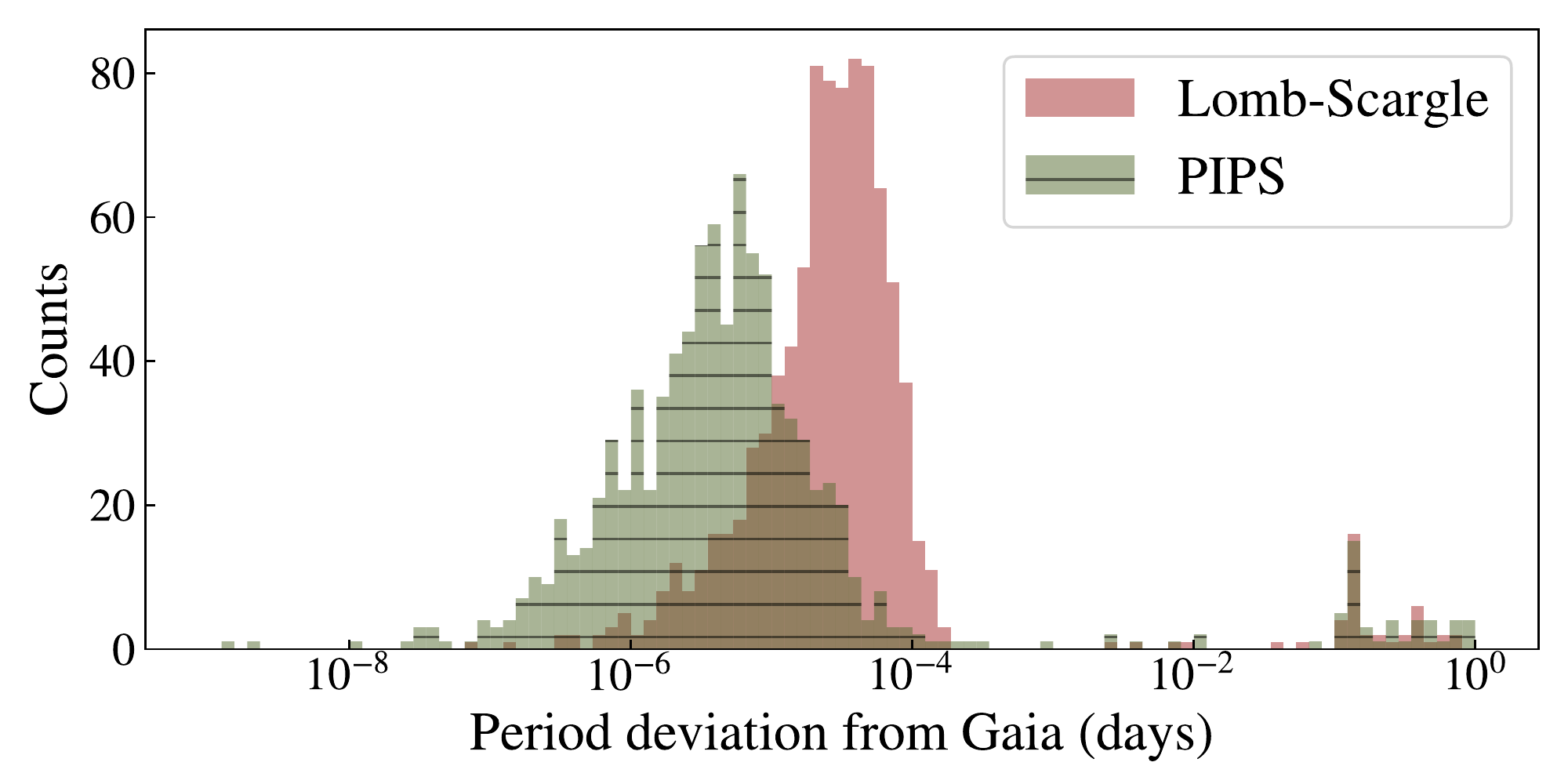}
                \caption{Deviation from Gaia period ($\Delta P$) when analysed with Lomb-Scargle (red) and \pipelinename{} (green).}
                \label{fig: hist_gaia_period}
            \end{figure}
            
            \begin{figure}
                \centering
                \includegraphics[width=\linewidth]{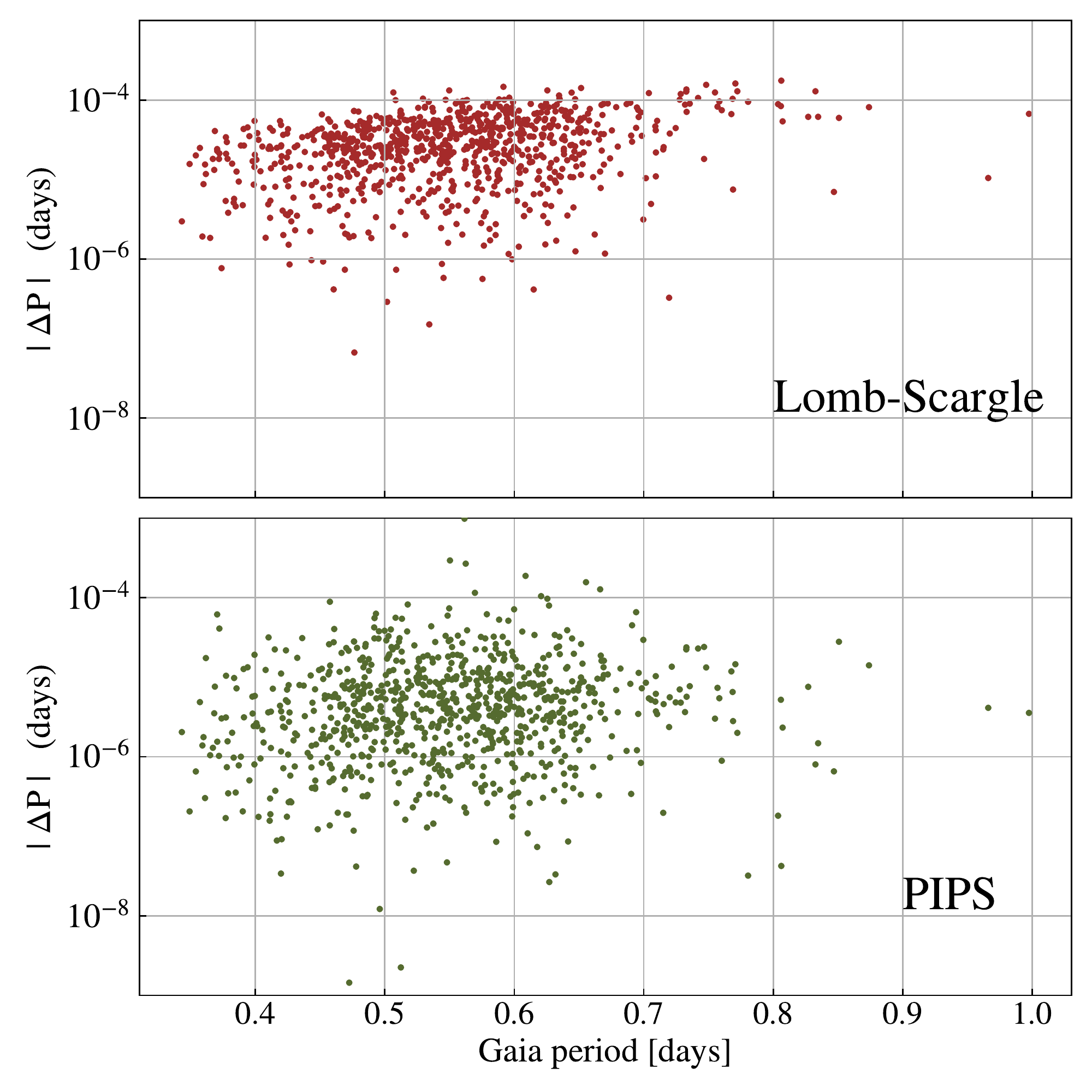}
                \caption{$|\Delta P|$ as a function of Gaia's period, for our stars with $\Delta P < 10^{-2}$ only. Lomb-Scargle shows a slight correlation between period and error, while our pipeline is more consistent across a large range of periods.}
                \label{fig:scatter_gaia_period}
            \end{figure}
            
            \begin{figure*}
            \centering
            \includegraphics[width=\linewidth]{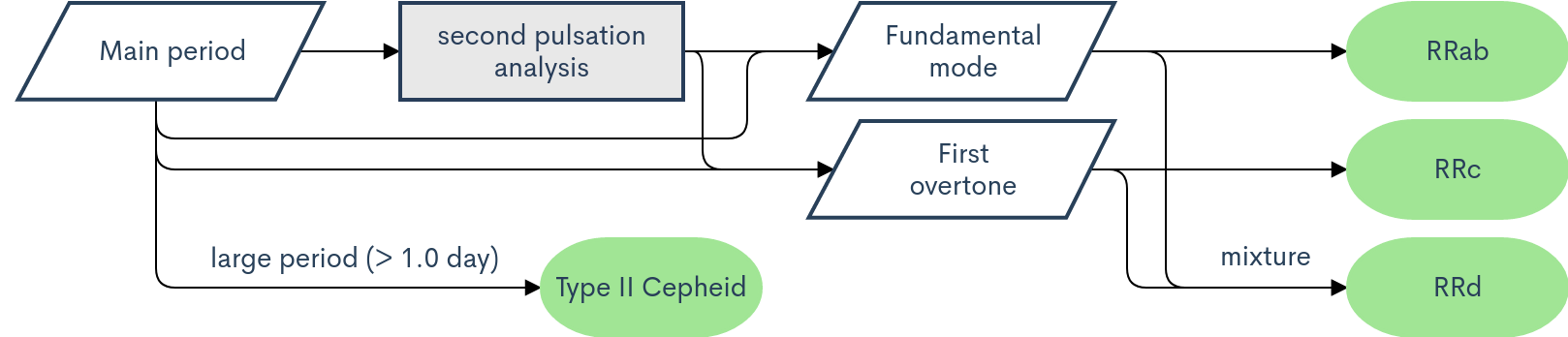}
            \caption{A flowchart showing the procedure to classify variable stars.}
            \label{fig: flowchart2 classification}
            \end{figure*}
        
            \subsubsection{Methodology}
            \label{sec: methodology}
            \pipelinename{} is a period-determination algorithm based on template-free, adaptive-function (``Free-Form'') fitting.
            The analysis begins with an initial guess from the Lomb-Scargle periodogram, but proceeds to refine the value and explore the associated uncertainty. We assume that when a light curve is folded about a more correct period, fitting will yield a smaller $\chi^2$ value compared to a folding about a value farther from the true pulsation period. We also assume that there will be one global minimum when examining the $\chi^2$ space, representing the true period of the star. The simplest way to find this correct period would be to iterate through the entirety of possible periods and perform a fit to the data folded with each period in turn. This is computationally intensive, inefficient, and complicates the exploration of uncertainty. We choose instead to provide an informed prior based on the Lomb-Scargle estimation, and use $\chi^2$ minimisation where both the period and the Fourier terms are free parameters. We find that this identifies the period more accurately and that this exploration of parameter space provides robust information regarding the uncertainty of the value. 

            Given time-series data, \pipelinename{} finds the best value for the principal pulsation period by checking $\chi^2$ values against the best-fit Fourier function which represents the analytic form of the light curve,
            \begin{equation}
                \label{eq: Fourier fitting}
                y_\mathrm{fit} = A_0 + \sum_{k=i}^{K_\mathrm{max}} \left[a_k \cos\left(\frac{2\pi k}{P}x\right) 
                        + b_k \sin\left(\frac{2\pi k}{P}x\right)\right].
            \end{equation}
            Here, $x$ is the phase-folded time data $x \equiv t\pmod P$, and $y_\mathrm{fit}$ represents the expected magnitude at that phase (the unitless phase is $x/P$). The parameters $A_0$, $a_k$, $b_k$, and  $P$, are determined by a linear-regression fitting using the \texttt{curve\_fit} function in \texttt{scipy} \citep{2020SciPy-NMeth}. In this analysis, we take $K_\mathrm{max}=5$, based on a cross-validation test using Gaia data of RR Lyrae stars.
            
            The number of parameters and intrinsic scatter (or a small SNR) may in some cases make it difficult to fit the correct value of $P$. This is primarily because Fourier parameters dominate the degrees of freedom, yielding many local minima in the $\chi^2$ space when viewed as a function of $P$. Additionally, the change in the $\chi^2$ value at different $P$ becomes less obvious as the SNR decreases, because even at the best $P$, the folded data exhibit a roughly correct shape but with large intrinsic scatter, yielding a relatively large $\chi^2$ value. These relations can be viewed as analogous to the mechanics of a potential well whose surface is the representation of the $\chi^2$ value. This well exhibits a larger friction force between sliding objects as the surface becomes rougher (many local minima) due to different combinations of Fourier coefficients. The SciPy \texttt{curve\_fit} function attempts to ``place an object'' (the initial guess) to perform tests with the goal of finding the ``bottom'' of the well. How far this test object slides down is a function of the slope. This ``slide test'' tends to be unsuccessful near the bottom of the potential, which can be seen in a nearly one-to-one relationship between the initial guess and the resulting ``best fit'' period. The ``initial kick'' from the linear regression algorithm yields a characteristic size in the scatter of the resulting period when compared to the initial guess values, and we take this as the statistical uncertainty of the period\footnote{Here, we use $\mathrm{std}(x)$ as a short notation for the standard deviation, $\mathrm{std}(x) = \sqrt{\left[\sum_i^N(x_i-\overline{x})\right]/N}$.}
            \begin{equation}
                \sigma_P = \mathrm{std}\left(P_\mathrm{fit} - P_\mathrm{trial}\right)
                            \bigg|_{P_\mathrm{trial}\approx P}.
            \end{equation}
            This process requires a fitting to Equation~\ref{eq: Fourier fitting} between $10^2$ and $10^4$ times, and our pipeline is designed to perform this analysis as fast as possible. When the uncertainty of the period is not required, it is sufficient and much faster to find the $P$ value which yields a minimum $\chi^2$ for Equation~\ref{eq: Fourier fitting} when evaluated as a function of $P$ (i.e., fixed $P$ within single fitting).

        \subsubsection{Performance Validation}
            
            We tested the period-detection function of our pipeline with raw RR Lyrae light curves from Gaia DR2 \citep{GaiaDR2_variability}. We chose a sample consisting of 1355 RR Lyrae stars (\texttt{gaiadr2.vari\_rrlyrae}). This set of data includes 910 RRab stars and 445 RRc or RRd stars, whose photometry in the $G$ band is taken for more than 30 epochs (\texttt{num\_clean\_epochs\_g} $> 30$) and is located nearby (\texttt{parallax} $> 0.25$) with less than 20\% uncertainty in its astrometry measurement (\texttt{parallax\_over\_error} $> 5$).
            We search for any period between 0.2 and 1.0\,d equally without an initial guess, with no visual inspection or human help.
            
            The results for this validation are shown in Figures~\ref{fig: hist_gaia_period}~and~\ref{fig:scatter_gaia_period}. In most cases both Lomb-Scargle (LS) and \pipelinename{} agreed with Gaia's period, although there is a population with $\sim 0.1$--1.0\,d error in period values. About 96\% of LS results and 95\% of \pipelinename{} results were considered ``good results,'' and \pipelinename{} outperforms LS with mean error an order of magnitude smaller than LS results. Moreover, LS exhibits a slight correlation between $\Delta P$ and period, an expected behaviour when a noninformed uniform prior in frequency space is used. This creates unequally-spaced windows in period space and thus the resolution becomes lower as the period increases. \pipelinename{} adaptively changes the search-window size, and therefore does not exhibit this issue.
            
            Since Lomb-Scargle is a widely accepted tool in the community, we consider our new package \pipelinename{} as an equally acceptable tool overall and one which may be better suited for our particular analysis based on these validation results.

    \subsection{Classification}
        \begin{figure*}
            \centering
            \includegraphics[width=\linewidth]{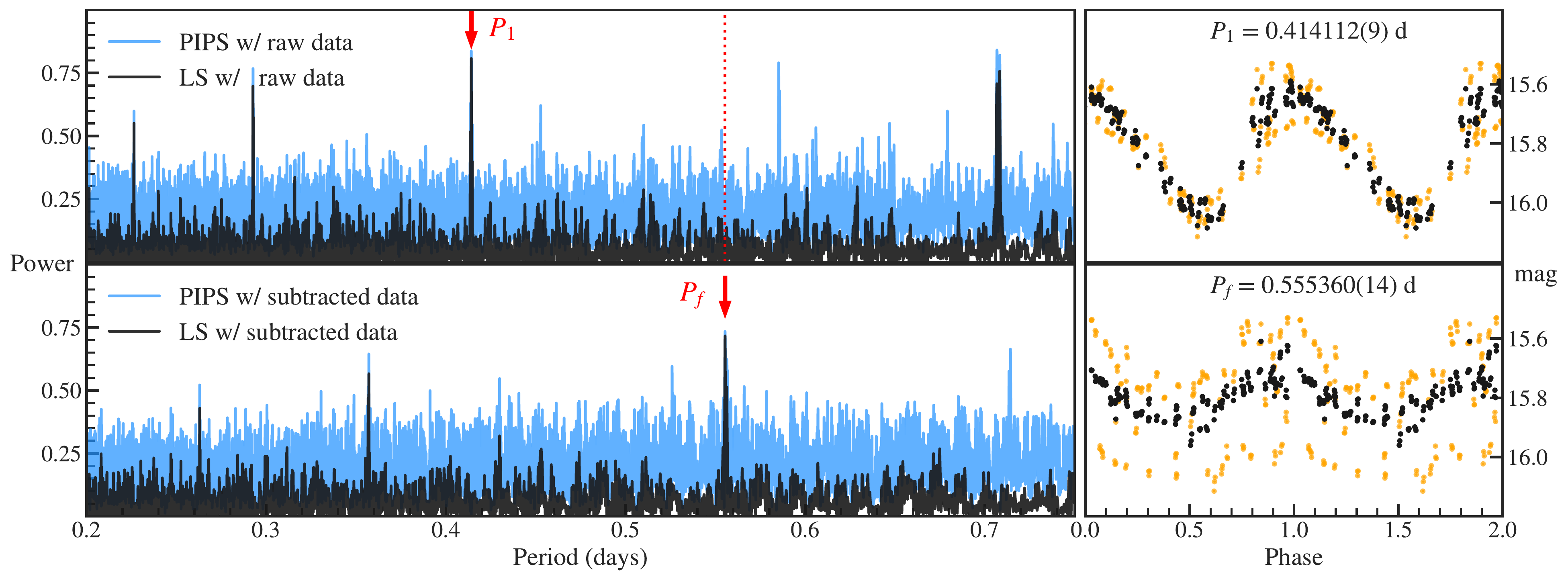}
            \caption{Left: Periodograms created for the intermediate steps in RRd classification. Blue (lighter color) represents the normalized periodogram created by PIPS and the black (darker color) represents the Lomb-Scargle periodogram. The power values in PIPS periodogram are compared with $\chi^2$ values evaluated against Eq.\ref{eq: Fourier fitting}, $\mathcal{P} = 1 - \chi^2/\chi^2_0$. Here, $\chi^2_0$ is the non-varying reference $\chi^2_0 = \sum (y_i-\overline{y})^2/\sigma_i^2$. Right: Decoupled modes at each detected period. Orange (lighter color) represents the raw data folded at each period, without subtraction. The abscissa represents the dimensionless phase (repeated twice).}
            \label{fig:RRd_periodogram}
        \end{figure*}

        RR Lyrae stars are generally classified based on their pulsation modes. This classification was first introduced by \cite{Bailey_1902} \revised{based on morphology}, and other than the consolidation of RRa and RRb stars into the RRab type, it has remained the accepted method. \revised{Although RR Lyrae stars as a group span a huge range of behaviors, within our sample we see that our }RR Lyrae types are largely separated by the fundamental mode pulsation, with RRab (also called RR0) having a 0.5--0.7\,d period and the first-overtone pulsation, RRc (RR1), having a $\sim 0.3$\,d period. Some stars exhibit a combination of these two modes, and are classified as RRd (RR01). The fundamental mode generally has a steep rise in magnitude, while the first overtone is closer to a sinusoid. The shapes of these light curves \revised{can be} explained using \revised{sophisticated hydrodynamical calculations} (see \cite{Stellingwerf1975}, \cite{Kollath2002}). \revisedII{Some RR Lyrae stars may exhibit significant modulations in amplitude, phase, and period over a few days \citep{JurcsikAJ132} to several years \citep{PrudilMNRAS466}.  This phenomenon is called the ``Blazhko effect" after the first observation \citep[see][]{Blazhko_1907}, and several attempts have been made to explain it \citep{SmolecBlazhko}. The more complex simultaneous modulations of amplitude and period are always present in Blazhko RRab stars \cite[See][]{Benko2010}. Modulations can also be seen in RRc stars, primarily (but not exclusively) in amplitude.}
        
        In order to classify the stars in our sample, we follow the general methodology seen in most RR Lyrae star and M15-specific studies \cite[e.g.,][]{Silbermann1995,Clementini_Gaia_2016}.
        Although the original method to distinguish between RRab and RRc stars is morphology-based and requires a careful statistical study of population and various parameters, it is now clear that RRab stars can be distinguished from RRc and RRd stars clearly in period space. This is especially true in M15, whose member objects share similar metallicities and ages \citep{APOGEE_M15_age_2020}. This prior knowledge about a narrow range of metallicities resolves the possible degeneracy that arises between different Oosterhoff groups \citep{Oosterhoff_1939}.
    
\begin{figure}
        \centering
        \includegraphics[width=\linewidth]{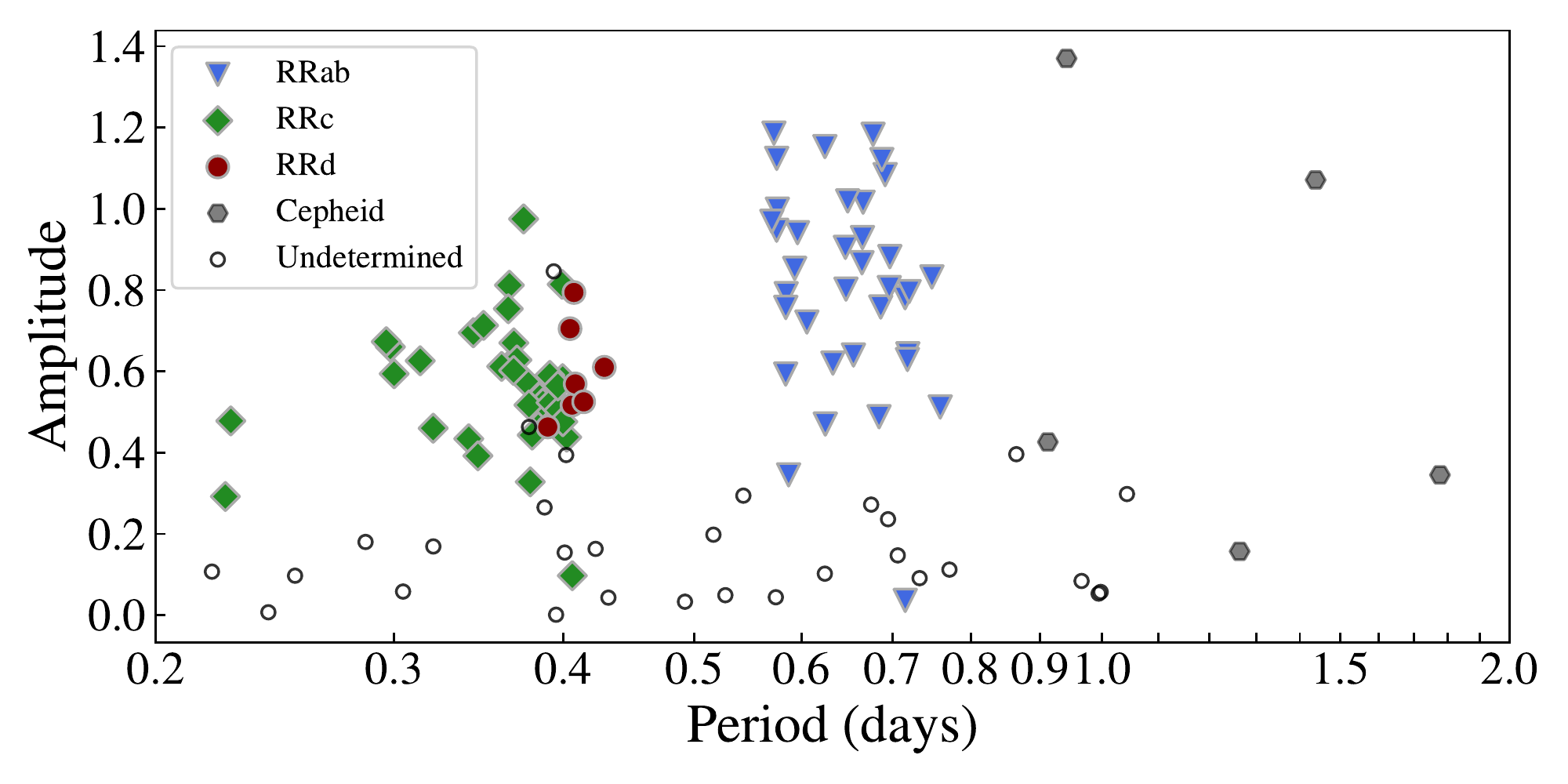}
        \caption{Classification based on the Bailey (period-amplitude) diagram. These data are used to make initial guesses for the dominant component in the double-mode analysis. RRd stars whose type is determined after double-mode analysis are overlayed.}
        \label{fig: period-amplitude}
    \end{figure}

        Our procedure to classify variable stars follows three steps:
        (1) based on existing studies, we confirm that the distribution of our detected periods agrees with those in the literature; 
        (2) we determine the separation between types, and
        (3) we confirm the overall agreement in various parameter spaces. Figure~\ref{fig: period-amplitude} shows the distribution of detected periods and amplitudes. As reported by \cite{Oosterhoff_1939} and \cite{Silbermann1995}, we see a gap between the two main populations at a period value of $\sim 0.5$\,d.
        
        Once the initial classification based on the main period is finished, we search for the secondary period to distinguish between RRc and RRd stars.
        \begin{figure}
            \centering
            \includegraphics[width=\linewidth]{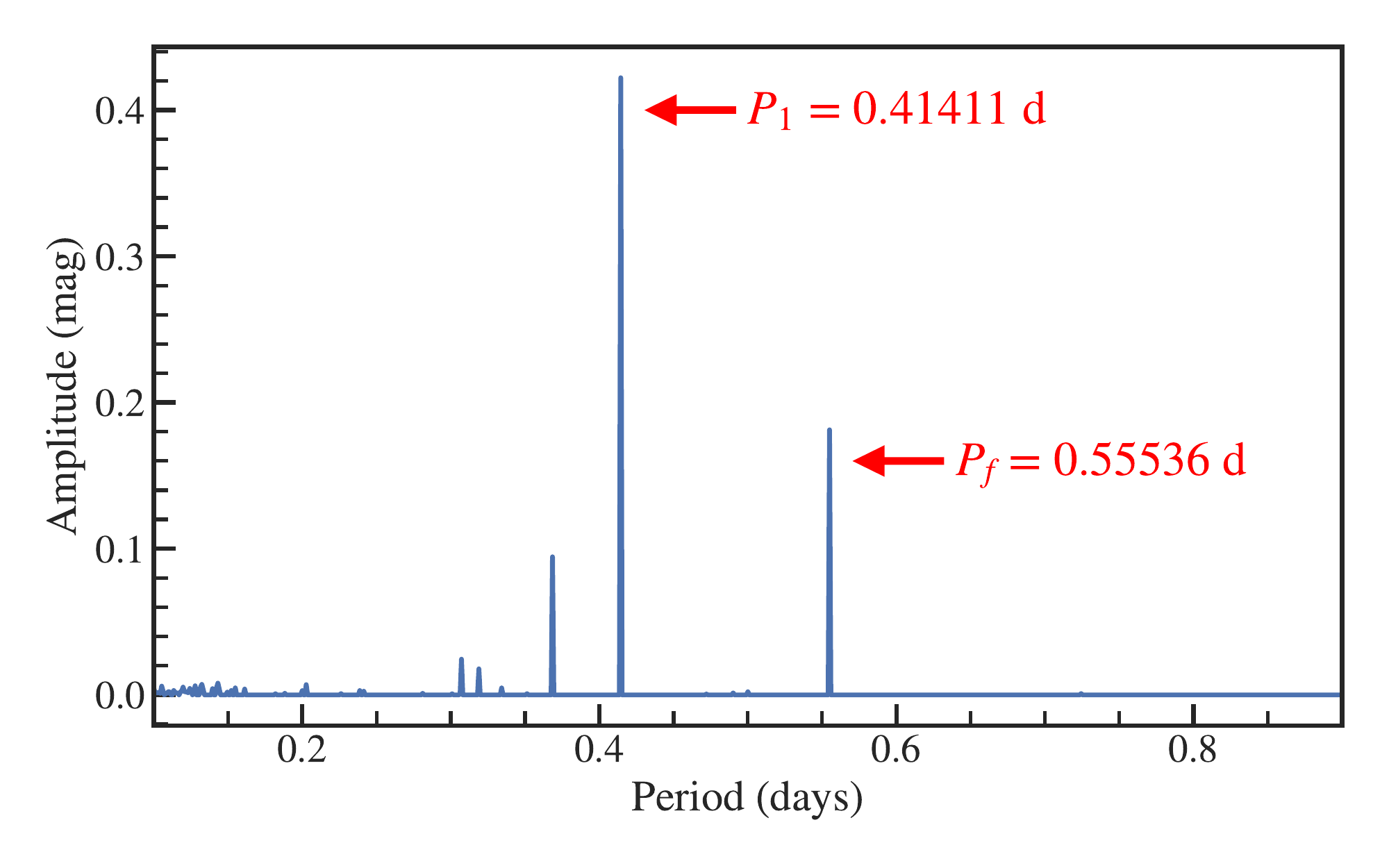}
            \caption{Amplitude Spectrum for V053 with our photometry data. The fundamental mode $P_1$ and the secondary pulsation (fundamental) mode $P_f$ can be clearly seen.}
            \label{fig:amplitude_periodogram}
        \end{figure}
        \revisedII{
        We employ a modified non-linear discrete Fourier transform (DFT) method to construct an amplitude spectrum. The basic fitting method has been used in previous studies of multi-mode RR Lyrae stars, such as \cite{Gruberbauer_2007_rrd}:
        \begin{equation}
            \text{mag} = A_0 + \sum_i^{i_\text{max}} A_i\cos\left(2\pi f_i t - \theta_i \right)\ .
        \end{equation}
        In order to further improve the accuracy of this method, we minimize the template effect as discussed in previous sections by fitting a template-free function (i.e., with higher harmonics):
        \begin{equation}
            \text{mag} = A_0 + \sum_i^{i_\text{max}} A_i\mathcal{F}_i\left(2\pi f_it\right)\ ,
        \end{equation}
        where $\mathcal{F}_i$ is the normalized best-fit Fourier series $\mathcal{F}_i(f) = \sum_k^{k_\text{max}} a_k \cos(2\pi fkt) + b_k \sin(2\pi fkt)$\footnote{Equivalently, $\mathcal{F}_i(f) = \sum_k^{k_\text{max}} c_k \cos(2\pi fkt - \theta_k)$.}.
        The frequency $f_i$ at which the function is evaluated is determined by our main period-search method described in Sec.~\ref{sec: methodology}. This method is effective at detecting the pulsation modes in the resulting amplitude spectrum, since the harmonic components of the target frequencies at each pulsation mode are fitted (i.e., removed) at each iteration. 
        Figure~\ref{fig:amplitude_periodogram} is an example of our obtained amplitude spectra. 
        }
        
        \revisedII{
        As our goal is to distinguish between RRc and RRd stars, this method can be further simplified to the analysis of the two strongest signals, $i=1,2$, for the first overtone and fundamental mode, respectively:
            \begin{equation}
                \label{eq: full doublemode fourier}
                V = A_0 + A_1\mathcal{F}_1\left(\frac{2\pi t}{P_1}\right)
                        + A_f\mathcal{F}_f\left(\frac{2\pi t}{P_f}\right)
            \end{equation}
        where $\mathcal{F}_1$ is the normalized best-fit Fourier function to the first overtone at $P_1$ and $\mathcal{F}_f$ is the Fourier function fitted to the fundamental pulsation mode at $P_f$. The method of searching for the secondary period $P_f$ is identical to searching for the main period as described in \ref{sec: methodology}, except that the $\chi^2$ value is evaluated against the residuals $V_\text{res,i} = y_i - A_1\mathcal{F}_1\left[(2\pi t_i)(P_1)\right]$ (`prewhitened' data, e.g. \citealt{Blomme2011}). This residual fitting is then further optimized with linear regression to fine-tune the value of $Pf$.
        This method is similar to that seen in e.g., \cite{Soszynski2009}, \cite{Jurksik_2015_RRd}, \cite{Moskalik2003}.
        }
        
        \revisedII{
        We identify RRd stars by first selecting stars with $0.7 < P_1/P_f < 0.8$ as suggested by \citep[e.g. ][]{Jurksik_2015_RRd}, and then evaluating the False Alarm Probability \citep[FAP; for review, see][]{VanderPlas_2018} of the detected secondary period in a Lomb-Scargle periodogram with a threshold of $\text{FAP} < 0.001$. This value is empirically determined to remove the most of the false secondary periods detected for RRab and RRc stars.
        The resulting periods and amplitudes signals are shown in Fig.~\ref{fig:amplitude_periodogram}. Using the two strongest signals, we can work to decouple the modes by subtracting each term $\mathcal{F}_i$ from the raw data. The intermediate periodogram used to select a peak to evaluate the amplitude and resulting decoupled modes is shown in Fig.~\ref{fig:RRd_periodogram}\footnote{The increased noise floor and signal level is an expected behaviour for a periodogram with multi-term fitting. For review, see \cite{VanderPlas_2018}.}, along with the resulting decoupled modes. 
        To check this procedure against known data, we performed the same analysis on OGLE RRd stars \citep{Soszynski2011_OGLE_BLG}. Our results are in a good agreement with the secondary period information provided by OGLE (for an example, see Fig.~\ref{fig:RRd_periodogram_OGLE}).
        }
        
\section{Results}
\label{sec: results}
    \subsection{Period Determinations}
        We have determined periods and types for \numstarsWithCeph{} variable stars, of which \numstars{} were RR Lyrae stars and five were Type II Cepheid stars. The results of our analysis are presented in Table~\ref{tab: main result table}, which includes the ID, position, best-fit period determination and its uncertainty, amplitude, and type classification for each star we considered. The period error is determined by taking the largest uncertainties out of three sources: instrumental (lowest digit of the telescope timestamp), statistical (randomness in the fitting algorithm), and algorithmic (insufficient $\chi^2$ search scaling) errors. We believe that the resulting uncertainties ($\sim10^{-7}$\,d) are more accurate and meaningful than many previously reported uncertainties, which are generally calculated by taking the instrumental timestamp uncertainty divided by the number of periods spanned during observations. We show our phase-folded light curves in Figures~\ref{fig: rrab}--\ref{fig: questionable}, with the addition of two decomposed pulsation modes for double-mode RRd stars.
        
        We compared our results to similar works in our references and found that there is general agreement between all works, although a small number of stars have classifications which vary between our references. Some of our results (e.g. V16, V35, V51, V93, V118) do not agree with the references, however, we note that our disagreement with the references is not more common than the disagreement between the references themselves.

\subsection{Stars by Classification Type}
    
        \begin{figure}
            \centering
            \includegraphics[width=\linewidth]{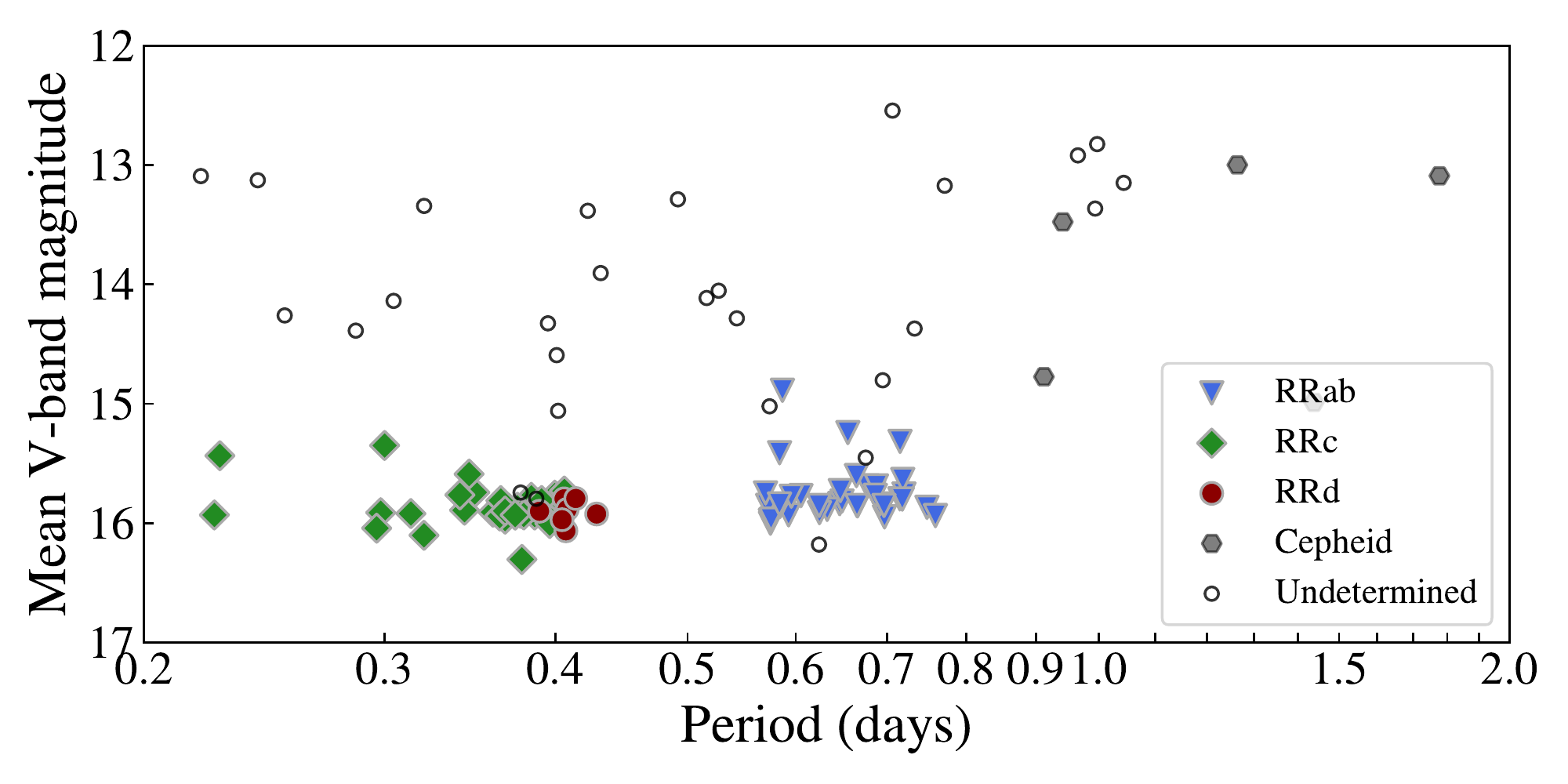}
            \caption{Mean $V$-band magnitude of star and main pulsation period of each star. Type II Cepheid variables are well isolated from RR Lyrae stars, and the brighter magnitude of undetermined stars suggests contamination from nearby stars.}
            \label{fig: period-magnitude}
        \end{figure}
        
        Out of 79 total variable stars, we find 33 RRab stars, with an average period of 0.65\,d. These stars are clearly distinguished by their sharply peaked maxima and are single-mode pulsators\footnote{Note that these distinguishing features are not used for classification, but that the clear differences support classification via separation in period-amplitude space.}. We classify \revisedII{34} stars as RRc with an average period of 0.35\,d, and with light curves showing near-sinusoidal shapes unique to first-overtone pulsation. We find \revisedII{7} RRd multimode stars with an average period of 0.41\,d, although we note that this period value is specific to the first overtone when considering multimode pulsators. \revised{With our double-mode analysis, we obtained mean value of $A_f/A_1$ of \revisedII{0.37} for RRd stars.} While less common in M15, we also detect 5 Type II Cepheid variables, with an average period of 1.26\,d. These Type II Cepheid light curves are similar to those of RRab stars, as they pulsate in the fundamental mode, but their forms vary and we do not perform a detailed analysis of them.

        \revised{As shown by \cite{Catelan2009_Oosterhoff}, Mean RRab periods and minimum RRab periods serve as reliable indicators of Oosterhoff type. Our value, $\langle P_{\mathrm{ab}}\rangle=0.65$ d and $(P_{\mathrm{ab}})_\mathrm{min}=0.57$ d suggests M15 is OoII type. This is consistent with previously reported results (e.g. \citealt{Oosterhoff_1939} and \citealt{Corwin_2008}).}

        We made 34 possible detections of variable stars, in addition to the \numstars{} about which we are confident. This undetermined population typically belongs to the ``large distance'' group identified in Figure~\ref{fig: distance to GCVS} and are not firmly classified owing to possible contamination and a small SNR. Stars marked with ``??'' are undetermined.
        
        The overall population distribution in the Bailey (period-amplitude) diagram (Fig. \ref{fig: period-amplitude}) shows some notable characteristics, such as (a) a short pulsation ($\sim 0.2$--0.4\,d) with a relatively small amplitude ($\lessapprox 0.9$\,mag) for RRc stars, (b) a narrow period range for RRd stars ($P\approx 0.39$--$0.43$\,d), (c) no confirmed detections of variable stars with $P \approx 0.5$\,d, and (d) RRab stars having a wide range in amplitude (0.4--1.3 mag) but a small range in period ($P \approx 0.55$--0.75\,d). These characteristics are consistent with \revised{existing studies of M15 (e.g. \cite{Corwin_2008}, \cite{Silbermann1995}, \cite{Ferro2006}). We note that RR Lyrae populations not constrained by the narrow metallicity range of M15 will display more variation}.
        
        
        When our stars are plotted in period-magnitude space as seen in Figure~\ref{fig: period-magnitude}, all types of RR Lyrae stars are tightly gathered at a similar apparent $V$-band magnitude of $V\approx 16$\,mag. This is consistent with previous observations \citep[e.g.,][]{Bingham1984,Silbermann1995}, and clearly shows the separation of Type II Cepheid stars from RR Lyrae stars, which are brighter and have longer pulsation times. Our undetermined stars are often brighter, and there is no consistency in their measured periods, which span values belonging to all RR Lyrae types as well as Type II Cepheids. This suggests that the majority of the undetermined stars are contaminated by nearby objects, raising their perceived brightness but leaving a wide range of period values.
        
        \revisedII{For stars with multi-mode pulsations, it is of interest to take the amplitude ratio of the fundamental mode, $A_f$, and the first overtone, $A_1$. In our sample of RRd stars we see a mean value of $\langle A_f/A_1\rangle \approx0.371$ (with a standard deviation of $\sigma_{Af/A1}\approx 0.065$), suggesting that this ratio is relatively consistent across our RRd stars. We note that our RRd stars have $A_f/A_1$ values $\le 1$, suggesting that the first overtone is slightly more dominant in these RRd stars. This is consistent with other observations, such as \cite{Jurksik_2015_RRd}.}
        
        The ratio of the periods of the fundamental mode and the first overtone is also a notable feature of the multimode RRd stars. Prior work suggests that this ratio should be quite narrow \citep[see][]{Clementini_Gaia_2016}, and our results support this with bounds of \revisedII{$0.745 \le P_1/P_f \le 0.747$ and a mean $\langle P_1/P_f \rangle = 0.746$}.
    
    \subsection{Notes on Individual Stars}

    \paragraph*{V027} -- We confirm that V27 does not appear be a variable star, as has been noticed since at least \cite{Hogg73}.
        
    \paragraph*{V095} -- This variable has a period of 1.263\,d, and peaks brighter than 13\,mag. No period determinations for this star can be found in the reference literature, likely owing to brightness contamination resulting from close proximity to the core of M15. The period and brightness of this star suggest that it is a Type II Cepheid variable.
    
    \paragraph*{V107} -- Both the light-curve shape and the period of 0.277\,d suggest that it is an RRc star. No previous determinations for the period can be found in the reference literature.
    \paragraph*{V123} -- While the light-curve shape is unclear, the period of 0.715\,d and the magnitude range suggest that this is an RRab star. No previous determinations for the period can be found in the reference literature.
    \paragraph*{V140} -- This variable peaks brighter than 13\,mag and has a relatively long period of 1.776\,d, suggesting that it is a Type II Cepheid variable. No previous determinations for the period can be found in the reference literature.
    
    \paragraph*{V155} -- Another bright star with a long period of 0.912\,d, this variable is also likely a Type II Cepheid, and does not have a previous period determination in the reference literature. 
    \paragraph*{Possible detections} -- Of those variables without classifications, we have detected 10 objects with no previously reported periods in our references:
        V079, V106a, V106b, V110, V115, V121, V143, V147, V150, V154, and V158. V106a shows a large secondary period with a ratio of $P_1/P_f\approx0.75$, suggesting it is a possible RRd star. These stars require further confirmation and investigation.
    
\section{Conclusion and future work}
\label{sec: long-term}
    Using the Nickel telescope at Lick Observatory, we observed M15 over \nickelnights{} nights and accumulated \nickelimgs{} images. In tandem, we have developed a custom pipeline to detect periodic behaviour in our data. We compared this new, purpose-built pipeline to existing techniques using the Gaia data release and found our results to be highly encouraging. We discussed the analysis of single and multimode pulsators, as well as the uncertainty treatment in our pipeline. Our analysis led to confident period classifications for \numstars{} RR Lyrae stars in M15, two of which do not have a value listed in the reference literature. Of these \numstars{} stars, the majority (67) are single-mode pulsators, which are classified as RRab and RRc stars. 
    
    Looking to the future, we note that the long-term period evolution of RR Lyrae stars is not well constrained by either theory or observation \citep{Jurcsik2017}. The difficulty of making a sufficiently accurate period determination, combined with the long timescales required to see this change, restrict the effectiveness of observationally constraining this behaviour, and theoretical derivations remain challenging \citep{Stellingwerf_overtone_1987}. Long-term, high-quality data are necessary for investigating the temporal evolution of RR Lyrae stars. With the addition of the results presented herein, we plan to conduct such an investigation in a subsequent paper (Murakami et al., in prep.).

\section*{Acknowledgements}
    We acknowledge Josh Bloom and Dan Weisz for thought-provoking discussions of period analysis and validation, and Arjun Savel for providing knowledge on data-analysis tools. 
    This research made use of Astropy,\footnote{http://www.astropy.org} a community-developed core Python package for astronomy \citep{astropy:2013, astropy:2018}. 
    We acknowledge generous support from the TABASGO Foundation, Marc J. Staley, the Christopher R. Redlich Fund, and the Miller Institute for Basic Research in Science (U.C. Berkeley). Research at Lick Observatory is partially supported by a generous gift from Google. We also thank the student observers involved with the research group: Samantha Stegman, Julia Hestenes, Keto Zhang, Raphael Baer-Way, Teagan Chapman, Matt Chu, Asia deGraw, Nachiket Girish, Romain Hardy, Connor Jennings, Evelyn Lu, Emily Ma, Emma McGinness, Shaunak Modak, Derek Perera, Druv Punjabi, Jackson Sipple, James Sunseri, Kevin Tang, Sergiy Vasylyev, Jeremy Wayland, and Abel Yagubyan. We also appreciate the assistance of the staff at Lick Observatory.
    
\section*{Data Availability}
    The raw photometry data used in our analysis is available at \url{https://github.com/SterlingYM/M15data2020}, and also upon request to an author.




\bibliographystyle{mnras}
\bibliography{ref.bib}
\vfill


\pagebreak
\onecolumn
\appendix

    \section{Table}

\begin{center}
\begin{longtable}{lllllllllll}

\caption{Results.}
\label{tab: main result table}

\\ \hline \hline
ID & $\alpha$(J2000)  & $\delta$(J2000) & $P$ & $\sigma_P$ & $P_2$ & $\sigma_{P2}$ & $A_1$ & $A_2$ & $\overline V$ & Type \\
& (deg) & (deg) & (days) & (days) & (days) & (days) & (mag) & (mag) & (mag) & \\
\endfirsthead

\multicolumn{11}{c}%
{{\tablename\ \thetable{} -- continued from previous page}} \\
\hline \hline
ID & $\alpha$(J2000)  & $\delta$(J2000) & $P$ & $\sigma_P$ & $P_2$ & $\sigma_{P2}$ & $A_1$ & $A_2$ & $\overline V$ & Type \\
& (deg) & (deg) & (days) & (days) & (days) & (days) & (mag) & (mag) & (mag) & \\
\hline 
\endhead

\hline \multicolumn{8}{r}{{Continued on next page}} \\ 
\endfoot

\hline \hline
\endlastfoot

        \hline
        V001  & 322.4590 & 12.1740 & 1.437837434 & 3.7e-07 &       -  &      -  & 1.07 &   -  &    14.99 &  CephII \\
        V002  & 322.4438 & 12.1687 & 0.684304375 & 1.6e-07 &       -  &      -  & 0.49 &   -  &    15.74 &    RRab \\
        V003  & 322.4222 & 12.1539 & 0.388720632 & 8.5e-06 &       -  &      -  & 0.48 &   -  &    15.83 &     RRc \\
        V004  & 322.4610 & 12.1216 & 0.313589117 & 2.4e-08 &       -  &      -  & 0.63 &   -  &    15.92 &     RRc \\
        V005  & 322.4646 & 12.1081 & 0.384211954 & 7.2e-08 &       -  &      -  & 0.54 &   -  &    15.78 &     RRc \\
        V006  & 322.4995 & 12.1886 & 0.666015094 & 1.7e-07 &       -  &      -  & 1.02 &   -  &    15.86 &    RRab \\
        V007  & 322.4955 & 12.1879 & 0.367563437 & 1.2e-06 &       -  &      -  & 0.63 &   -  &    15.97 &     RRc \\
        V008  & 322.4924 & 12.2027 & 0.646238777 & 3.9e-06 &       -  &      -  & 0.91 &   -  &    15.84 &    RRab \\
        V009  & 322.4967 & 12.2059 & 0.715282097 & 5.7e-07 &       -  &      -  & 0.78 &   -  &    15.80 &    RRab \\
        V010  & 322.5284 & 12.1680 & 0.386389721 & 2.1e-06 &       -  &      -  & 0.52 &   -  &    15.93 &     RRc \\
        V011  & 322.5416 & 12.1617 & 0.343265510 & 5.5e-08 &       -  &      -  & 0.69 &   -  &    15.89 &     RRc \\
        V012  & 322.5387 & 12.1535 & 0.592875644 & 5.2e-07 &       -  &      -  & 0.85 &   -  &    15.93 &    RRab \\
        V013  & 322.5288 & 12.1485 & 0.574910937 & 1.4e-07 &       -  &      -  & 0.95 &   -  &    15.98 &    RRab \\
        V015  & 322.5164 & 12.0831 & 0.584394482 & 1.6e-07 &       -  &      -  & 0.79 &   -  &    15.86 &    RRab \\
        V016  & 322.5210 & 12.2035 & 0.399195555 & 3.3e-07 &       -  &      -  & 0.81 &   -  &    15.84 &     RRc \\
        V017  & 322.5163 & 12.1981 & 0.428912296 & 1.5e-07 & 0.575678 & 1.3e-05 & 0.61 & 0.20 &    15.92 &     RRd \\
        V018  & 322.5145 & 12.1954 & 0.367725969 & 1.0e-07 &       -  &      -  & 0.67 &   -  &    15.87 &     RRc \\
        V019  & 322.5240 & 12.2121 & 0.572306432 & 1.9e-06 &       -  &      -  & 1.19 &   -  &    15.85 &    RRab \\
        V020  & 322.5156 & 12.1648 & 0.696958720 & 4.3e-07 &       -  &      -  & 0.88 &   -  &    15.95 &    RRab \\
        V021  & 322.5023 & 12.1513 & 0.648795261 & 3.6e-06 &       -  &      -  & 1.02 &   -  &    15.82 &    RRab \\
        V022  & 322.3988 & 12.1539 & 0.720230076 & 1.6e-06 &       -  &      -  & 0.80 &   -  &    15.76 &    RRab \\
        V023  & 322.5466 & 12.2388 & 0.632698496 & 1.6e-07 &       -  &      -  & 0.62 &   -  &    15.89 &    RRab \\
        V024  & 322.4624 & 12.1654 & 0.369691742 & 1.1e-07 &       -  &      -  & 0.63 &   -  &    15.87 &     RRc \\
        V025  & 322.5786 & 12.1650 & 0.665318041 & 6.2e-06 &       -  &      -  & 0.93 &   -  &    15.85 &    RRab \\
        V026  & 322.4986 & 12.2595 & 0.402317102 & 4.4e-08 &       -  &      -  & 0.44 &   -  &    15.94 &     RRc \\
        V029  & 322.5385 & 12.2263 & 0.575015865 & 1.5e-05 &       -  &      -  & 1.12 &   -  &    16.00 &    RRab \\
        V030  & 322.4458 & 12.1661 & 0.405998183 & 8.0e-08 & 0.544739 & 1.3e-05 & 0.52 & 0.19 &    15.80 &     RRd \\
        V031  & 322.4602 & 12.2352 & 0.408183108 & 6.0e-08 & 0.547051 & 1.9e-05 & 0.57 & 0.21 &    15.88 &     RRd \\
        V032  & 322.4781 & 12.1972 & 0.605303420 & 1.3e-07 &       -  &      -  & 0.72 &   -  &    15.77 &    RRab \\
        V033  & 322.4812 & 12.1593 & 0.583940164 & 3.5e-08 &       -  &      -  & 0.76 &   -  &    15.41 &    RRab \\
        V034  & 322.4771 & 12.1521 & 0.400960355 &      -  &       -  &      -  & 0.15 &   -  &       -  &      ?? \\
        V035  & 322.4834 & 12.1219 & 0.624546598 & 2.6e-07 &       -  &      -  & 0.47 &   -  &    15.91 &    RRab \\
        V036  & 322.4850 & 12.1447 & 0.624130821 & 2.2e-06 &       -  &      -  & 1.15 &   -  &    15.85 &    RRab \\
        V038  & 322.4950 & 12.1268 & 0.375280597 & 4.2e-08 &       -  &      -  & 0.57 &   -  &    15.87 &     RRc \\
        V039  & 322.4986 & 12.1328 & 0.389551691 & 5.2e-08 & 0.522278 & 1.2e-05 & 0.46 & 0.16 &    15.90 &     RRd \\
        V040  & 322.5301 & 12.1351 & 0.377331483 & 4.1e-08 &       -  &      -  & 0.52 &   -  &    15.92 &     RRc \\
        V041  & 322.5107 & 12.1521 & 0.391761515 & 2.5e-06 &       -  &      -  & 0.52 &   -  &    15.88 &     RRc \\
        V042  & 322.5575 & 12.1575 & 0.360188851 & 5.2e-08 &       -  &      -  & 0.61 &   -  &    15.91 &     RRc \\
        V044  & 322.5185 & 12.1685 & 0.595669649 & 4.8e-06 &       -  &      -  & 0.94 &   -  &    15.77 &    RRab \\
        V045  & 322.5116 & 12.1588 & 0.677404150 & 1.9e-06 &       -  &      -  & 1.18 &   -  &    15.69 &    RRab \\
        V046  & 322.5090 & 12.1764 & 0.691443810 & 3.6e-06 &       -  &      -  & 1.08 &   -  &    15.84 &    RRab \\
        V047  & 322.5052 & 12.1664 & 0.687547509 & 1.4e-06 &       -  &      -  & 1.12 &   -  &    15.69 &    RRab \\
        V048  & 322.5093 & 12.2092 & 0.364972072 & 7.2e-08 &       -  &      -  & 0.81 &   -  &    15.81 &     RRc \\
        V049  & 322.5038 & 12.2136 & 0.655186807 & 4.3e-06 &       -  &      -  & 0.64 &   -  &    15.24 &    RRab \\
        V050  & 322.5391 & 12.1954 & 0.298060627 & 4.6e-08 &       -  &      -  & 0.66 &   -  &    15.91 &     RRc \\
        V051  & 322.4942 & 12.1928 & 0.396962319 & 1.3e-07 &       -  &      -  & 0.50 &   -  &    15.93 &     RRc \\
        V052  & 322.5473 & 12.1616 & 0.575654124 & 3.6e-06 &       -  &      -  & 1.00 &   -  &    15.97 &    RRab \\
        V053  & 322.4667 & 12.1365 & 0.414109598 & 9.3e-08 & 0.555360 & 1.4e-05 & 0.52 & 0.18 &    15.79 &     RRd \\
        V054  & 322.4956 & 12.1919 & 0.399572367 & 6.1e-08 &       -  &      -  & 0.58 &   -  &    15.95 &     RRc \\
        V055  & 322.5114 & 12.1622 & 0.748671788 & 3.9e-07 &       -  &      -  & 0.83 &   -  &    15.87 &    RRab \\
        V056  & 322.5089 & 12.1676 & 0.570260593 & 1.7e-07 &       -  &      -  & 0.97 &   -  &    15.75 &    RRab \\
        V057  & 322.5141 & 12.1522 & 0.349277148 & 5.6e-08 &       -  &      -  & 0.71 &   -  &    15.74 &     RRc \\
        V058  & 322.4770 & 12.1697 & 0.407285449 & 9.1e-08 & 0.545823 & 1.1e-05 & 0.79 & 0.41 &    16.06 &     RRd \\
        V059  & 322.5040 & 12.1769 & 0.431889913 &      -  &       -  &      -  & 0.04 &   -  &       -  &      ?? \\
        V060  & 322.5080 & 12.1510 & 0.718689553 & 3.6e-07 &       -  &      -  & 0.64 &   -  &    15.63 &    RRab \\
        V061  & 322.4737 & 12.1558 & 0.399688532 & 9.9e-08 &       -  &      -  & 0.48 &   -  &    15.77 &     RRc \\
        V062  & 322.4724 & 12.1780 & 0.377318789 &      -  &       -  &      -  & 0.46 &   -  &       -  &      ?? \\
        V063  & 322.5064 & 12.1760 & 0.646891427 & 1.4e-07 &       -  &      -  & 0.80 &   -  &    15.73 &    RRab \\
        V064  & 322.4795 & 12.1728 & 0.364249404 & 1.2e-07 &       -  &      -  & 0.75 &   -  &    15.94 &     RRc \\
        V065  & 322.4637 & 12.1566 & 0.718199490 & 4.5e-06 &       -  &      -  & 0.63 &   -  &    15.80 &    RRab \\
        V066  & 322.4735 & 12.1362 & 0.379343054 & 3.7e-08 &       -  &      -  & 0.44 &   -  &    15.93 &     RRc \\
        V067  & 322.4682 & 12.1644 & 0.404610074 & 1.2e-07 & 0.542406 & 1.0e-05 & 0.71 & 0.24 &    15.97 &     RRd \\
        V068  & 322.4830 & 12.1703 & 0.387438607 &      -  &       -  &      -  & 0.26 &   -  &       -  &      ?? \\
        V069  & 322.4822 & 12.1607 & 0.586772955 & 3.0e-07 &       -  &      -  & 0.35 &   -  &    14.89 &    RRab \\
        V070  & 322.4831 & 12.1620 & 0.367582599 & 7.4e-08 &       -  &      -  & 0.60 &   -  &    15.90 &     RRc \\
        V071  & 322.4827 & 12.1631 & 0.105774416 &      -  &       -  &      -  & 0.14 &   -  &       -  &      ?? \\
        V072  & 322.4923 & 12.1769 & 0.686283385 & 2.8e-07 &       -  &      -  & 0.76 &   -  &    15.77 &    RRab \\
        V073  & 322.4907 & 12.1724 & 0.401962763 &      -  &       -  &      -  & 0.39 &   -  &       -  &      ?? \\
        V074  & 322.5035 & 12.1436 & 0.296010243 & 4.6e-08 &       -  &      -  & 0.67 &   -  &    16.04 &     RRc \\
        V075  & 322.4927 & 12.1578 & 0.526924552 &      -  &       -  &      -  & 0.05 &   -  &       -  &      ?? \\
        V076  & 322.4926 & 12.1610 & 0.320665464 &      -  &       -  &      -  & 0.17 &   -  &       -  &      ?? \\
        V077  & 322.4898 & 12.1621 & 0.706426874 &      -  &       -  &      -  & 0.15 &   -  &       -  &      ?? \\
        V078  & 322.4906 & 12.1806 & 0.664751545 & 1.9e-07 &       -  &      -  & 0.87 &   -  &    15.60 &    RRab \\
        V079  & 322.4986 & 12.1611 & 0.285756395 &      -  &       -  &      -  & 0.18 &   -  &       -  &      ?? \\
        V081  & 322.4877 & 12.1650 & 0.253480390 &      -  &       -  &      -  & 0.10 &   -  &       -  &      ?? \\
        V082  & 322.4879 & 12.1678 & 0.491929862 &      -  &       -  &      -  & 0.03 &   -  &       -  &      ?? \\
        V083  & 322.4982 & 12.1661 & 0.516328627 &      -  &       -  &      -  & 0.20 &   -  &       -  &      ?? \\
        V084  & 322.4996 & 12.1634 & 0.543305073 &      -  &       -  &      -  & 0.29 &   -  &       -  &      ?? \\
        V086  & 322.4964 & 12.1686 & 0.941440619 & 5.5e-07 &       -  &      -  & 1.37 &   -  &    13.48 &  CephII \\
        V087  & 322.5010 & 12.1613 & 0.771371647 &      -  &       -  &      -  & 0.11 &   -  &       -  &      ?? \\
        V089  & 322.4851 & 12.1666 & 0.304592260 &      -  &       -  &      -  & 0.06 &   -  &       -  &      ?? \\
        V090  & 322.5025 & 12.1682 & 0.151975380 &      -  &       -  &      -  & 0.68 &   -  &       -  &      ?? \\
        V091  & 322.5110 & 12.1753 & 0.390929100 & 1.2e-07 &       -  &      -  & 0.59 &   -  &    15.81 &     RRc \\
        V092  & 322.4957 & 12.1603 & 0.373886851 & 8.7e-08 &       -  &      -  & 0.97 &   -  &    15.93 &     RRc \\
        V093  & 322.5003 & 12.1587 & 0.340598884 & 6.8e-08 &       -  &      -  & 0.43 &   -  &    15.76 &     RRc \\
        V094  & 322.4946 & 12.1750 & 0.395125307 &      -  &       -  &      -  & 0.00 &   -  &       -  &      ?? \\
        V095  & 322.4939 & 12.1559 & 1.263577492 & 8.5e-07 &       -  &      -  & 0.16 &   -  &    13.00 &  CephII \\
        V096  & 322.5393 & 12.2273 & 0.396360363 & 9.5e-08 &       -  &      -  & 0.56 &   -  &    16.00 &     RRc \\
        V097  & 322.4701 & 12.1753 & 0.696354571 & 9.1e-07 &       -  &      -  & 0.81 &   -  &    15.85 &    RRab \\
        V098  & 322.4735 & 12.1800 & 0.624063736 &      -  &       -  &      -  & 0.10 &   -  &       -  &      ?? \\
        V099  & 322.5013 & 12.2209 & 0.225141886 & 4.6e-08 &       -  &      -  & 0.29 &   -  &    15.93 &     RRc \\
        V100  & 322.4969 & 12.1566 & 0.320595991 & 8.0e-08 &       -  &      -  & 0.46 &   -  &    16.10 &     RRc \\
        V102  & 322.5129 & 12.1755 & 0.759381107 & 1.9e-06 &       -  &      -  & 0.51 &   -  &    15.94 &    RRab \\
        V103  & 322.4217 & 12.0908 & 0.583845056 & 1.4e-06 &       -  &      -  & 0.59 &   -  &    15.84 &    RRab \\
        V106a & 322.4846 & 12.1701 & 0.393562714 &      -  &       -  &      -  & 0.85 &   -  &       -  &      ?? \\
        V106b & 322.4840 & 12.1716 & 0.997568069 &      -  &       -  &      -  & 0.06 &   -  &       -  &      ?? \\
        V107  & 322.4842 & 12.1604 & 0.227240473 & 3.5e-08 &       -  &      -  & 0.48 &   -  &    15.43 &     RRc \\
        V110  & 322.5022 & 12.1562 & 0.675195633 &      -  &       -  &      -  & 0.27 &   -  &       -  &      ?? \\
        V111  & 322.5048 & 12.1674 & 0.378130163 & 7.5e-08 &       -  &      -  & 0.33 &   -  &    16.30 &     RRc \\
        V113  & 322.4947 & 12.0980 & 0.406180567 & 8.3e-08 &       -  &      -  & 0.10 &   -  &    15.73 &     RRc \\
        V114  & 322.4935 & 12.1777 & 0.345933528 & 9.3e-08 &       -  &      -  & 0.39 &   -  &    15.59 &     RRc \\
        V115  & 322.4962 & 12.1647 & 2.712286292 &      -  &       -  &      -  & 0.21 &   -  &       -  &      ?? \\
        V116  & 322.4970 & 12.1532 & 0.965689926 &      -  &       -  &      -  & 0.08 &   -  &       -  &      ?? \\
        V117  & 322.4992 & 12.1571 & 0.994196708 &      -  &       -  &      -  & 0.05 &   -  &       -  &      ?? \\
        V118  & 322.4981 & 12.1821 & 0.299995772 & 7.8e-08 &       -  &      -  & 0.59 &   -  &    15.35 &     RRc \\
        V119  & 322.4975 & 12.1697 & 0.733184352 &      -  &       -  &      -  & 0.09 &   -  &       -  &      ?? \\
        V121  & 322.4840 & 12.2093 & 0.242234009 &      -  &       -  &      -  & 0.01 &   -  &       -  &      ?? \\
        V122  & 322.5660 & 12.1741 & 0.574149201 &      -  &       -  &      -  & 0.04 &   -  &       -  &      ?? \\
        V123  & 322.4289 & 12.1648 & 0.715498035 & 4.3e-06 &       -  &      -  & 0.04 &   -  &    15.32 &    RRab \\
        V129  & 322.4922 & 12.1629 & 0.220068365 &      -  &       -  &      -  & 0.11 &   -  &       -  &      ?? \\
        V140  & 322.4943 & 12.1673 & 1.776476681 & 1.1e-06 &       -  &      -  & 0.35 &   -  &    13.09 &  CephII \\
        V143  & 322.4946 & 12.1612 & 0.694855607 &      -  &       -  &      -  & 0.24 &   -  &       -  &      ?? \\
        V145  & 322.4973 & 12.1674 & 0.422570740 &      -  &       -  &      -  & 0.16 &   -  &       -  &      ?? \\
        V147  & 322.4958 & 12.1657 & 3.072484243 &      -  &       -  &      -  & 0.11 &   -  &       -  &      ?? \\
        V150  & 322.4950 & 12.1715 & 1.043200501 &      -  &       -  &      -  & 0.30 &   -  &       -  &      ?? \\
        V154  & 322.4916 & 12.1708 & 0.864274136 &      -  &       -  &      -  & 0.40 &   -  &       -  &      ?? \\
        V155  & 322.4907 & 12.1697 & 0.911891427 & 2.7e-07 &       -  &      -  & 0.43 &   -  &    14.77 &  CephII \\
        V158  & 322.4676 & 12.1031 & 0.118082890 &      -  &       -  &      -  & 0.68 &   -  &       -  &      ?? \\

\end{longtable}
\end{center}

    
    \begin{center}
\begin{longtable}{llrrrrrrrr}

\caption{Calibration stars used with LPP$^*$.}
\label{tab: calibration stars}

\\ \hline \hline
\# &  Field &    $\alpha_\mathrm{obs}$ & $\delta_\mathrm{obs}$ &   $\alpha_\mathrm{diff}$ &  $\delta_\mathrm{diff}$ & $V_\mathrm{cal}$ & $V_\mathrm{obs}$ & $V_\mathrm{diff}$ &   $\sigma V_\mathrm{obs}$\\
 & & (deg) & (deg) & (deg) & (deg) & (mag) & (mag) & (mag) & (mag) \\
\endfirsthead

\multicolumn{8}{c}%
{{\tablename\ \thetable{} -- continued from previous page}} \\
\hline \hline
\# &  Field &    $\alpha_\mathrm{obs}$ & $\delta_\mathrm{obs}$ &   $\alpha_\mathrm{diff}$ &  $\delta_\mathrm{diff}$ & $V_\mathrm{cal}$ & $V_\mathrm{obs}$ & $V_\mathrm{diff}$ &   $\sigma V_\mathrm{obs}$\\
 & & (deg) & (deg) & (deg) & (deg) & (mag) & (mag) & (mag) & (mag) \\
\hline 
\endhead

\hline \multicolumn{8}{r}{{Continued on next page}} \\ 
\endfoot

\hline \hline
\endlastfoot

       \hline
        0  &  M15\_1 &  322.519265 &  12.127833 &  0.000105 &  0.000066 &  15.273728 &  15.287983 &  0.014255 &  0.014855 \\
        1  &  M15\_1 &  322.561284 &  12.140017 &  0.000107 &  0.000079 &  15.464006 &  15.442372 & -0.021634 &  0.012730 \\
        2  &  M15\_1 &  322.523107 &  12.118028 &  0.000107 &  0.000066 &  15.506678 &  15.530928 &  0.024250 &  0.012090 \\
        3  &  M15\_1 &  322.533446 &  12.085643 &  0.000082 &  0.000070 &  15.992274 &  15.963761 & -0.028513 &  0.013641 \\
        4  &  M15\_1 &  322.543151 &  12.125756 &  0.000090 &  0.000072 &  15.990031 &  15.979094 & -0.010937 &  0.014745 \\
        5  &  M15\_1 &  322.534900 &  12.136032 &  0.000092 &  0.000070 &  16.048878 &  16.052094 &  0.003216 &  0.013421 \\
        6  &  M15\_1 &  322.530015 &  12.109165 &  0.000103 &  0.000065 &  16.478746 &  16.496761 &  0.018014 &  0.019075 \\
        7  &  M15\_1 &  322.554640 &  12.141452 &  0.000097 &  0.000077 &  16.790747 &  16.790150 & -0.000597 &  0.020650 \\
        8  &  M15\_1 &  322.567823 &  12.134789 &  0.000100 &  0.000079 &  17.252762 &  17.252872 &  0.000110 &  0.031925 \\
        9  &  M15\_2 &  322.459429 &  12.150704 &  0.000109 &  0.000056 &  14.273025 &  14.260324 & -0.012701 &  0.013764 \\
        10 &  M15\_2 &  322.458902 &  12.159175 &  0.000105 &  0.000055 &  14.919133 &  14.922602 &  0.003468 &  0.010524 \\
        11 &  M15\_2 &  322.445463 &  12.140651 &  0.000109 &  0.000055 &  14.930785 &  14.917324 & -0.013461 &  0.009058 \\
        12 &  M15\_2 &  322.456988 &  12.156429 &  0.000100 &  0.000061 &  15.089972 &  15.072546 & -0.017426 &  0.015132 \\
        13 &  M15\_2 &  322.458802 &  12.131101 &  0.000093 &  0.000046 &  15.189565 &  15.185824 & -0.003741 &  0.016033 \\
        14 &  M15\_2 &  322.447355 &  12.151231 &  0.000096 &  0.000063 &  15.412107 &  15.412935 &  0.000828 &  0.010156 \\
        15 &  M15\_2 &  322.451633 &  12.155373 &  0.000094 &  0.000066 &  15.419053 &  15.430990 &  0.011937 &  0.022853 \\
        16 &  M15\_2 &  322.462334 &  12.144905 &  0.000083 &  0.000051 &  15.454053 &  15.469824 &  0.015771 &  0.030308 \\
        17 &  M15\_2 &  322.455935 &  12.140771 &  0.000115 &  0.000051 &  15.496215 &  15.507713 &  0.011498 &  0.018345 \\
        18 &  M15\_3 &  322.521463 &  12.222239 &  0.000095 &  0.000064 &  14.395865 &  14.393613 & -0.002252 &  0.014736 \\
        19 &  M15\_3 &  322.519897 &  12.196303 &  0.000098 &  0.000067 &  14.553714 &  14.588363 &  0.034649 &  0.014014 \\
        20 &  M15\_3 &  322.543955 &  12.236424 &  0.000120 &  0.000077 &  15.516245 &  15.496363 & -0.019882 &  0.026960 \\
        21 &  M15\_3 &  322.548378 &  12.234880 &  0.000116 &  0.000082 &  15.678872 &  15.663113 & -0.015759 &  0.034030 \\
        22 &  M15\_3 &  322.521121 &  12.227391 &  0.000099 &  0.000068 &  15.904249 &  15.909863 &  0.005614 &  0.014118 \\
        23 &  M15\_3 &  322.533092 &  12.203466 &  0.000118 &  0.000068 &  15.965824 &  15.989238 &  0.023414 &  0.027297 \\
        24 &  M15\_3 &  322.536526 &  12.208844 &  0.000107 &  0.000068 &  15.991126 &  15.992488 &  0.001362 &  0.021195 \\
        25 &  M15\_3 &  322.540593 &  12.240000 &  0.000117 &  0.000073 &  16.044006 &  16.013488 & -0.030518 &  0.022373 \\
        26 &  M15\_4 &  322.448676 &  12.204609 &  0.000102 &  0.000045 &  14.526760 &  14.510128 & -0.016632 &  0.015407 \\
        27 &  M15\_4 &  322.448944 &  12.191813 &  0.000106 &  0.000057 &  15.115295 &  15.136328 &  0.021032 &  0.014781 \\
        28 &  M15\_4 &  322.445159 &  12.195775 &  0.000119 &  0.000055 &  15.254565 &  15.249278 & -0.005288 &  0.014280 \\
        29 &  M15\_4 &  322.457367 &  12.208246 &  0.000083 &  0.000051 &  15.447569 &  15.461878 &  0.014309 &  0.013821 \\
        30 &  M15\_4 &  322.435704 &  12.192173 &  0.000132 &  0.000054 &  15.380196 &  15.364628 & -0.015568 &  0.023735 \\
        31 &  M15\_4 &  322.464925 &  12.212959 &  0.000096 &  0.000056 &  15.770203 &  15.753778 & -0.016425 &  0.020545 \\
        32 &  M15\_4 &  322.469867 &  12.195119 &  0.000086 &  0.000062 &  15.896890 &  15.904128 &  0.007238 &  0.024449 \\
        33 &  M15\_4 &  322.437433 &  12.198128 &  0.000116 &  0.000048 &  16.264444 &  16.270178 &  0.005733 &  0.024551 \\
        34 &  M15\_4 &  322.456705 &  12.210754 &  0.000098 &  0.000048 &  15.987055 &  15.979578 & -0.007477 &  0.018576 \\
        35 &  M15\_4 &  322.457820 &  12.194107 &  0.000090 &  0.000065 &  16.547810 &  16.548878 &  0.001068 &  0.022561 \\
        
\end{longtable}

$^*${The magnitude values are all in the $V$ band.\\ ``diff'' is
the deviation of the observed value (``obs'') from PS1 data (``cal'').}
\end{center}

    \clearpage

    \section{Lightcurves} 
        \begin{figure}
            \centering
            \includegraphics[width=\linewidth,page=1]{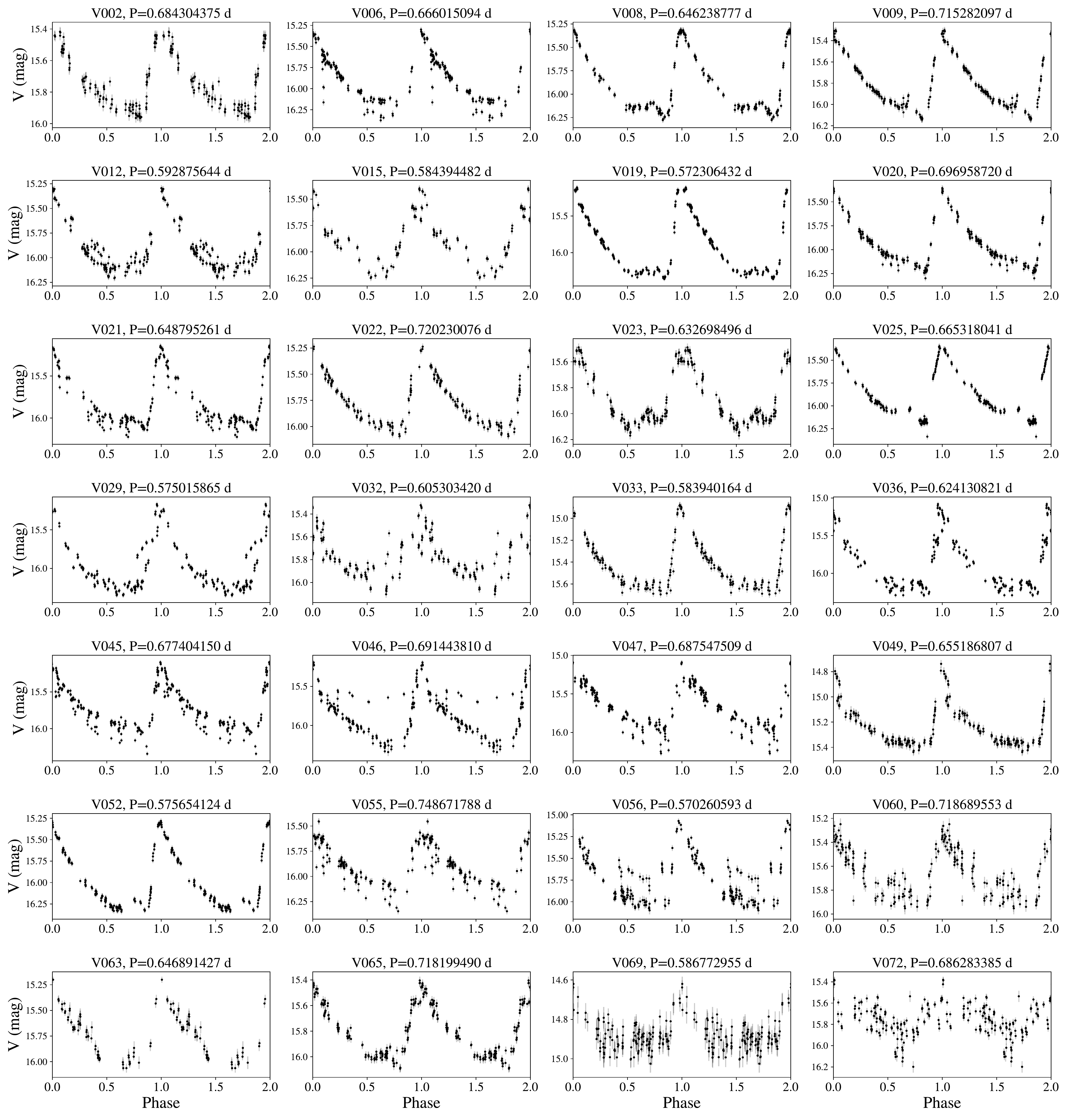}
            \caption{RRab stars: each cell is a folded light curve of a single-mode star pulsating in the fundamental mode (RRab). The abscissa is phase (normalised to a single period and repeated twice), and the ordinate is observed $V$-band magnitude.}
            \label{fig: rrab}
        \end{figure}
        
        \begin{figure}
            \centering
            \includegraphics[width=\linewidth,page=2,trim={0cm 37cm 0cm 0cm}]{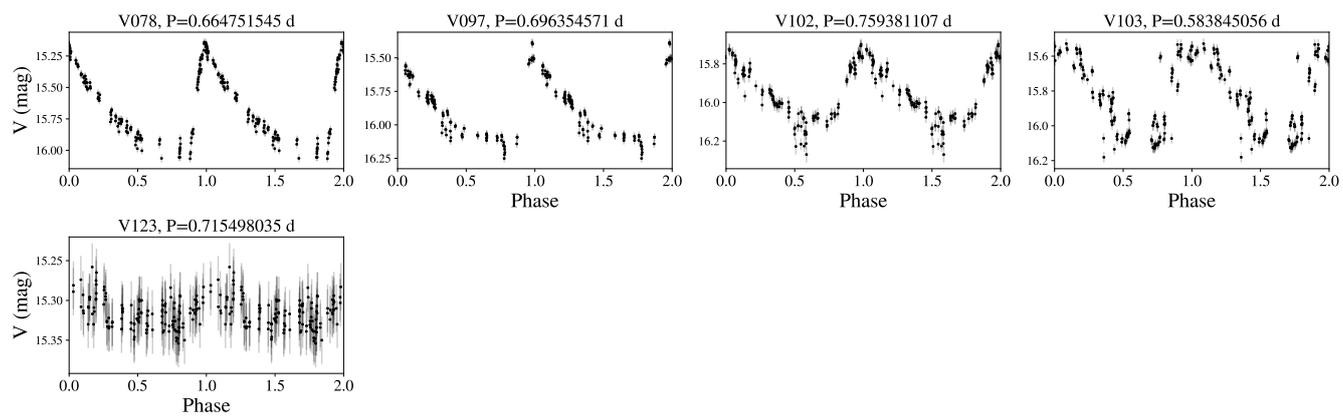}
            \caption{RRab stars (continued).}
            \label{fig: rrab2}
        \end{figure}
        
        \begin{figure}
            \centering
            \includegraphics[width=\linewidth,page=1]{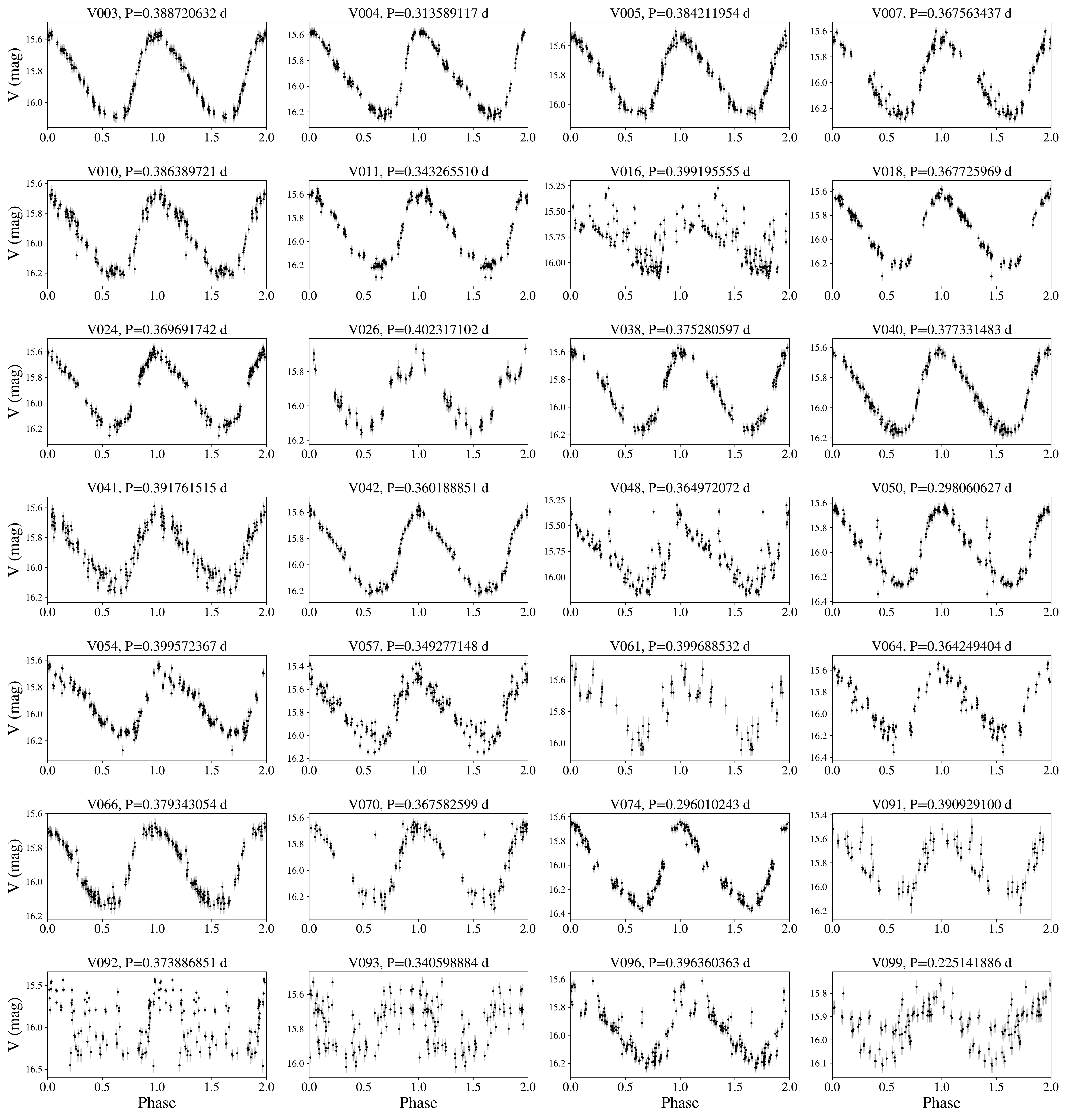}
            \caption{RRc stars. Conventions are the same as in Figure~\ref{fig: rrab}.}
            \label{fig: rrc}
        \end{figure}

        \begin{figure}
            \centering
            \includegraphics[width=\linewidth,page=2,trim={0cm 37.5cm 0cm 0cm}]{RRc_revII.pdf}
            \caption{RRc stars (continued).}
            \label{fig: rrc2}
        \end{figure}
        
        \begin{figure}
            \centering
            \includegraphics[width=\linewidth]{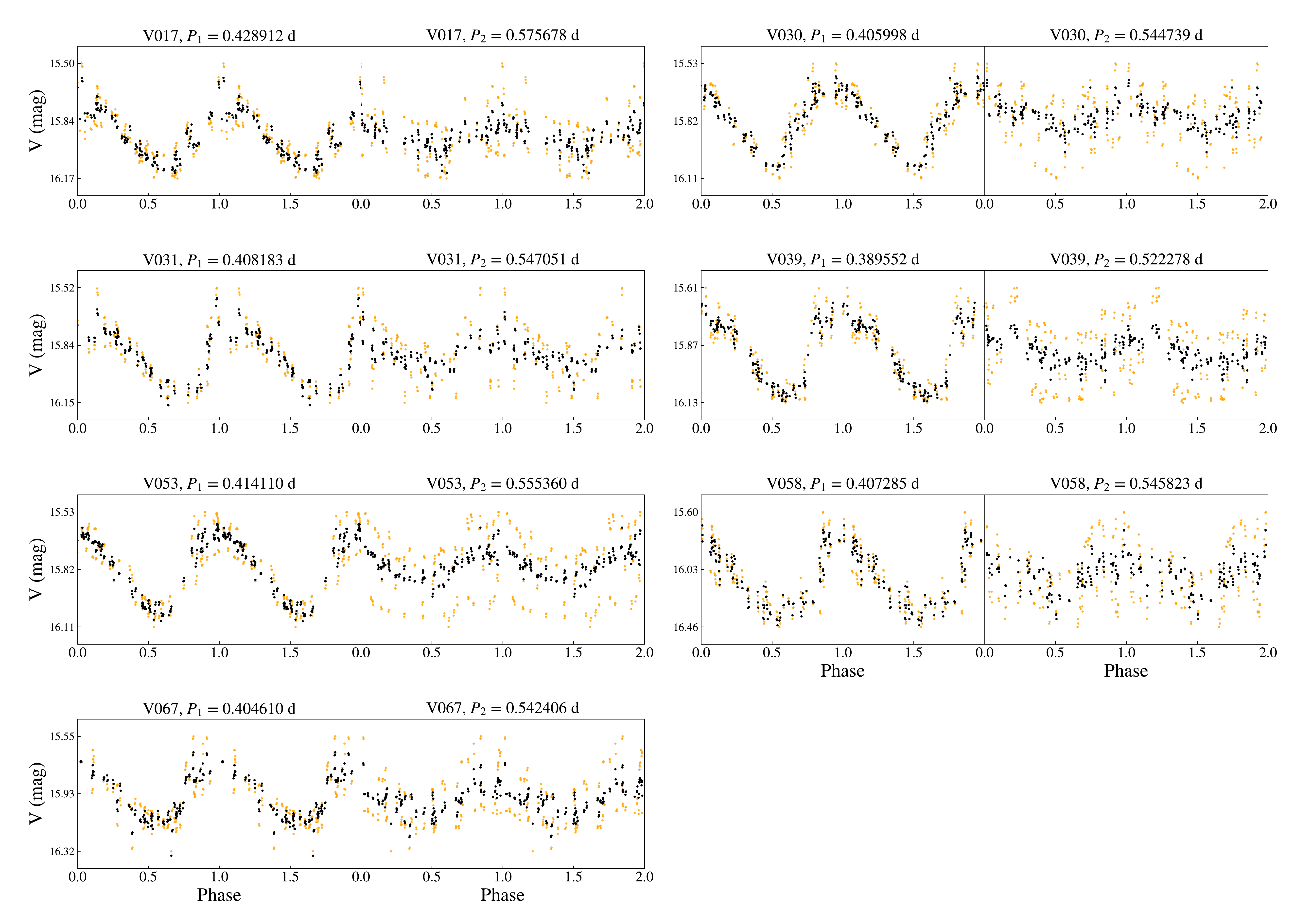}
            \caption{RRd stars: each set of plots represents a star pulsating in two periods simultaneously (RRd). 
            Left columns show raw (yellow, thin color) and decoupled (black) data folded at the first overtone period ($P_1$). Similarly, right columns show raw and decoupled data folded at the fundamental mode ($P_f$).}
            \label{fig: rrd1}
        \end{figure}

        \begin{figure}
            \centering
            \includegraphics[width=\linewidth,trim={0cm 37.5cm 0cm 0cm}]{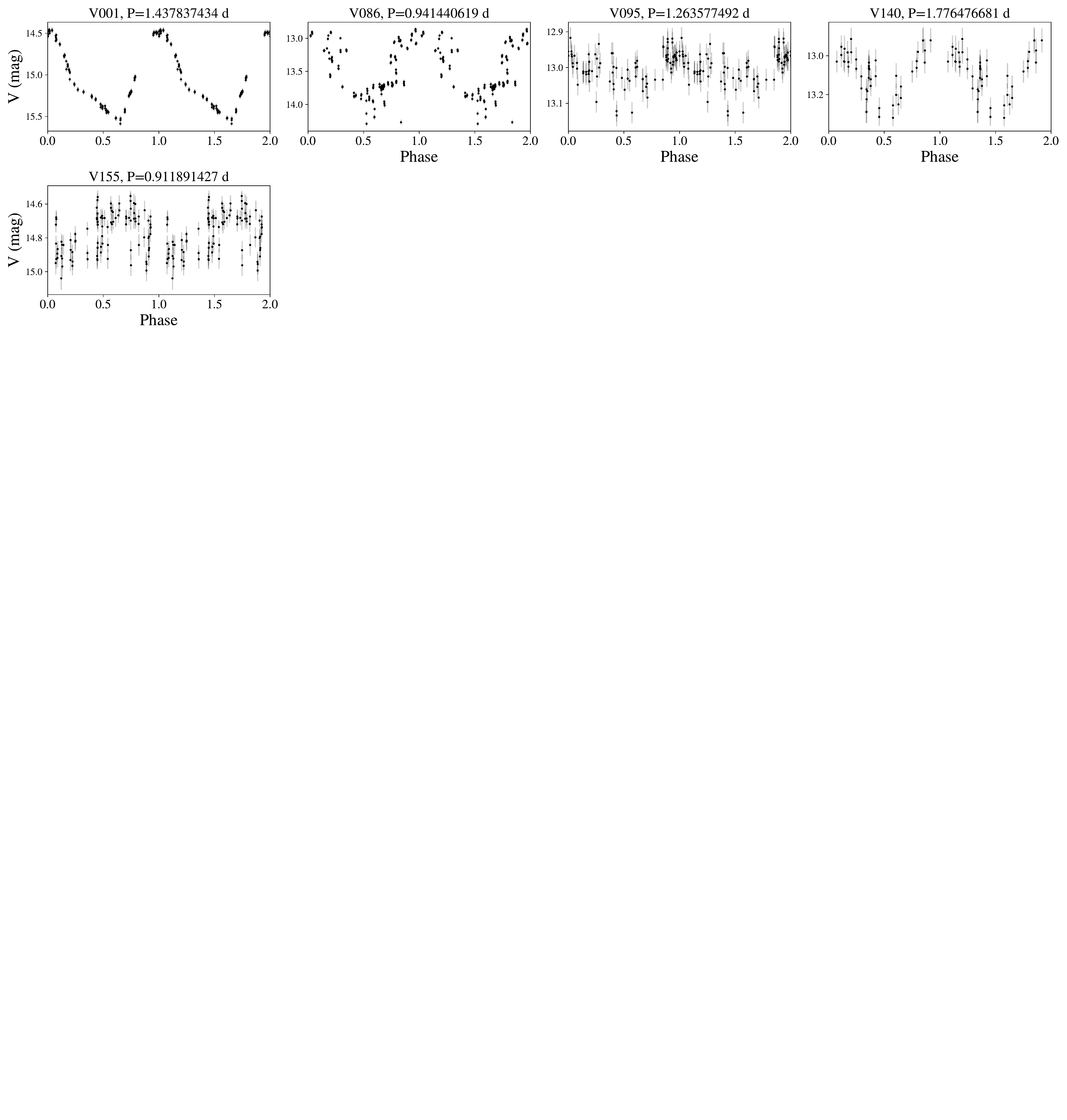}
            \caption{Type II Cepheid stars: each cell is a folded light curve of a single-mode star pulsating in the fundamental mode. While these stars exhibit morphological characteristics similar to those of RRab stars, these stars are brighter and have larger period values. The abscissa is phase (normalised to a single period and repeated twice), and the ordinate is observed $V$-band magnitude.}
            \label{fig:cepheid}
        \end{figure}

        \begin{figure}
            \centering
            \includegraphics[width=\linewidth,page=1]{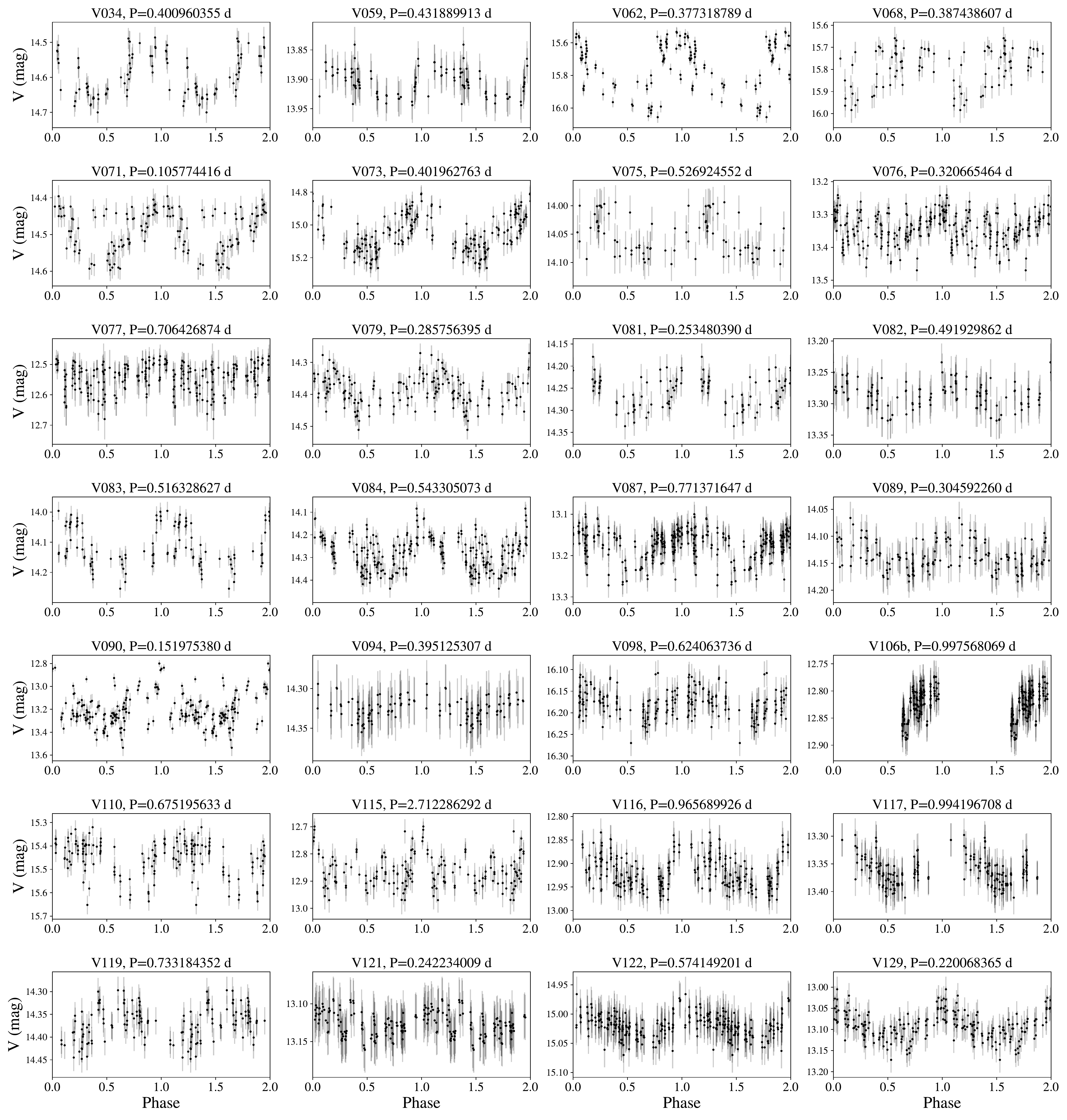}
            \caption{Stars with possible contamination (``undetermined'' stars). Some of our detections had relatively large distances from known coordinates. This resulted in possible contamination by nearby stars and small SNR with relatively large intrinsic scatter. These stars are treated as questionable, and are not processed for type classification, although some light curves indicate that they may be RRc stars by their morphology and main period. RRab stars and short-period Type II Cepheid stars cannot be distinguished owing to their common pulsation mode and inaccurate apparent magnitude. Since many of these stars do not have previously reported period values, we suggest our period estimation here.}
            \label{fig: questionable}
        \end{figure}

        \begin{figure}
            \centering
            \includegraphics[width=\linewidth,page=2,trim={0cm 37cm 0cm 0cm}]{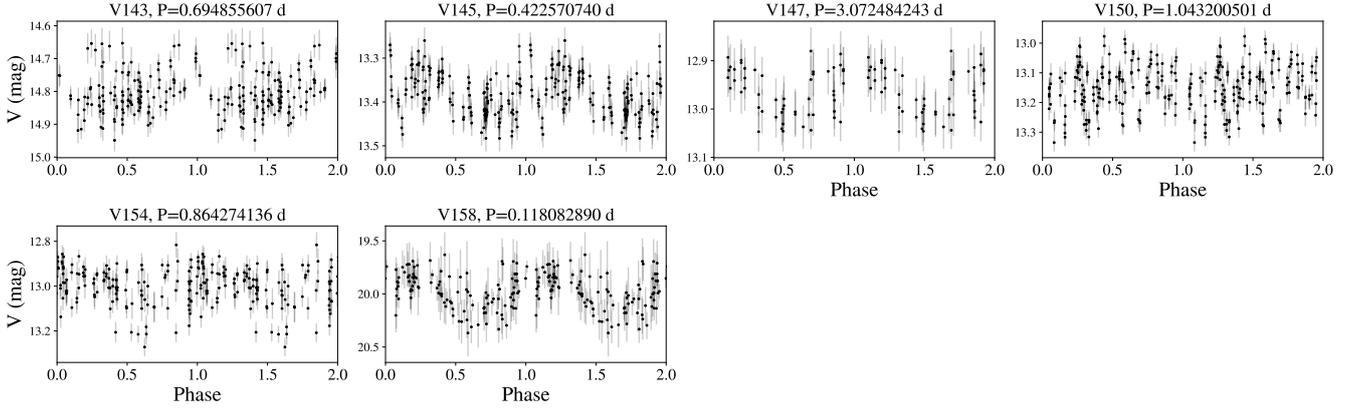}
            \caption{Stars with possible contamination (``undetermined'' stars, continued).}
            \label{fig: questionable2}
        \end{figure}

        \begin{figure}
            \centering
            \includegraphics[width=\linewidth]{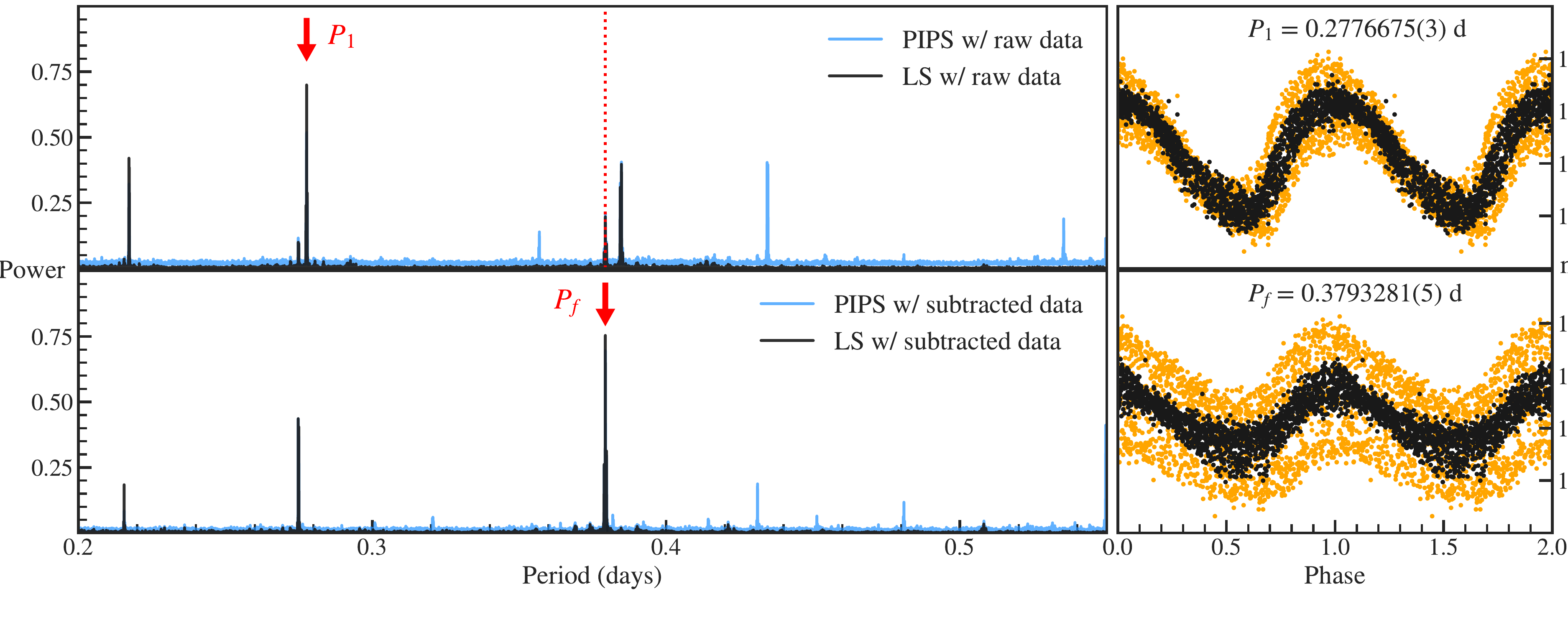}
            \caption{Same as Fig.~\ref{fig:RRd_periodogram}, but for an RRd star reported by OGLE-III \protect\cite[OGLE-BLG-RRLYR-09258;][]{Soszynski2011_OGLE_BLG} as an example to validate our method. Our detection of both periods is in \textit{perfect} agreement with the reported values by OGLE, $P_{1,\mathrm{OGLE}}=0.2776675(1)$ and $P_{f,\mathrm{OGLE}}=0.3793281(3)$.}
            \label{fig:RRd_periodogram_OGLE}
        \end{figure}
        
        \begin{figure}
            \centering
            \includegraphics[width=\linewidth]{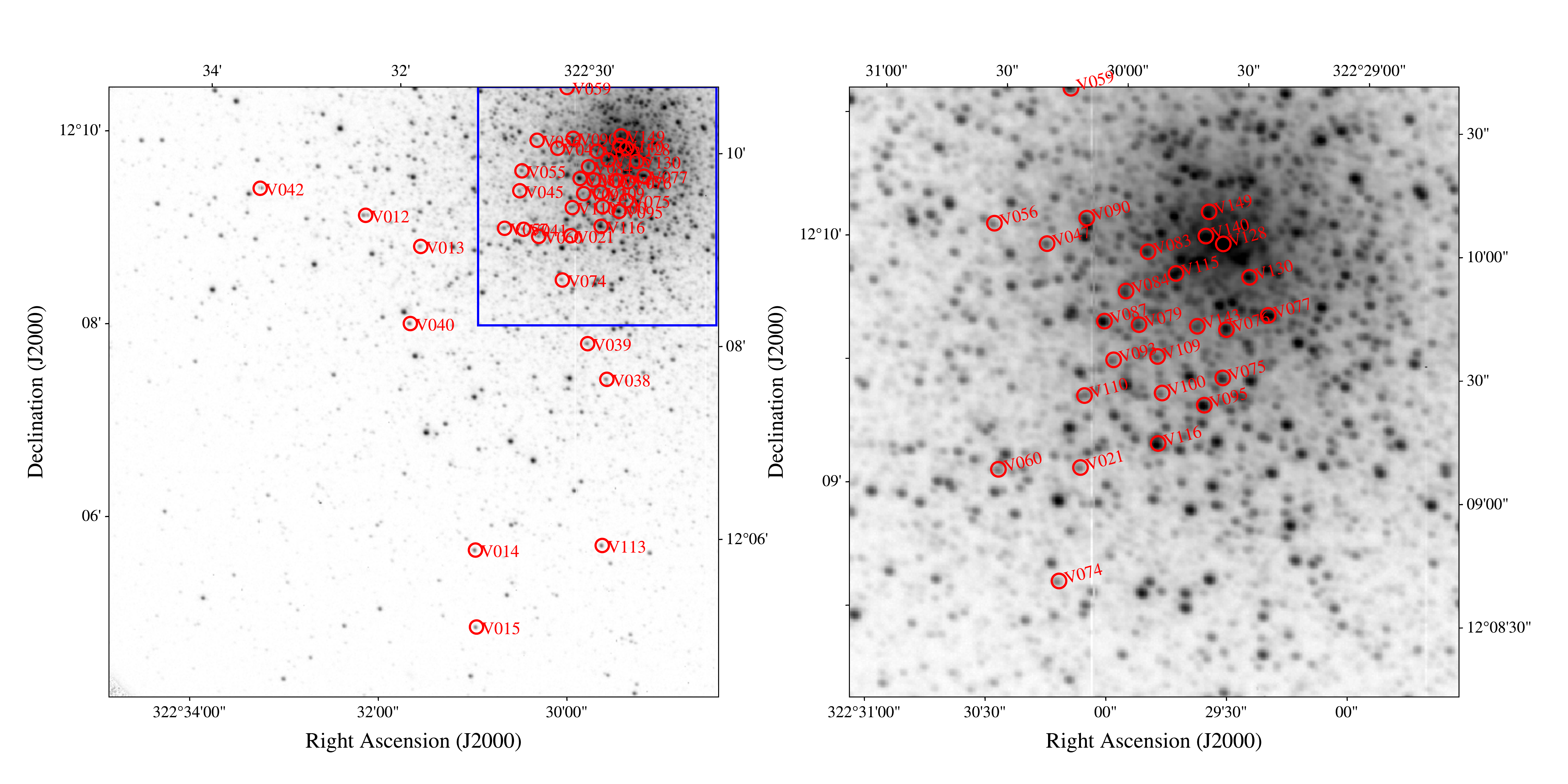}
            \caption{Left: $V$ band image of the field M15\_1 and stars whose data from this field is used for our final results. Blue rectangle indicates the area shown in the right. Right: A close-up image near the centre.}
            \label{fig: M15_1_coord}
        \end{figure}
        
        \begin{figure}
            \centering
            \includegraphics[width=\linewidth]{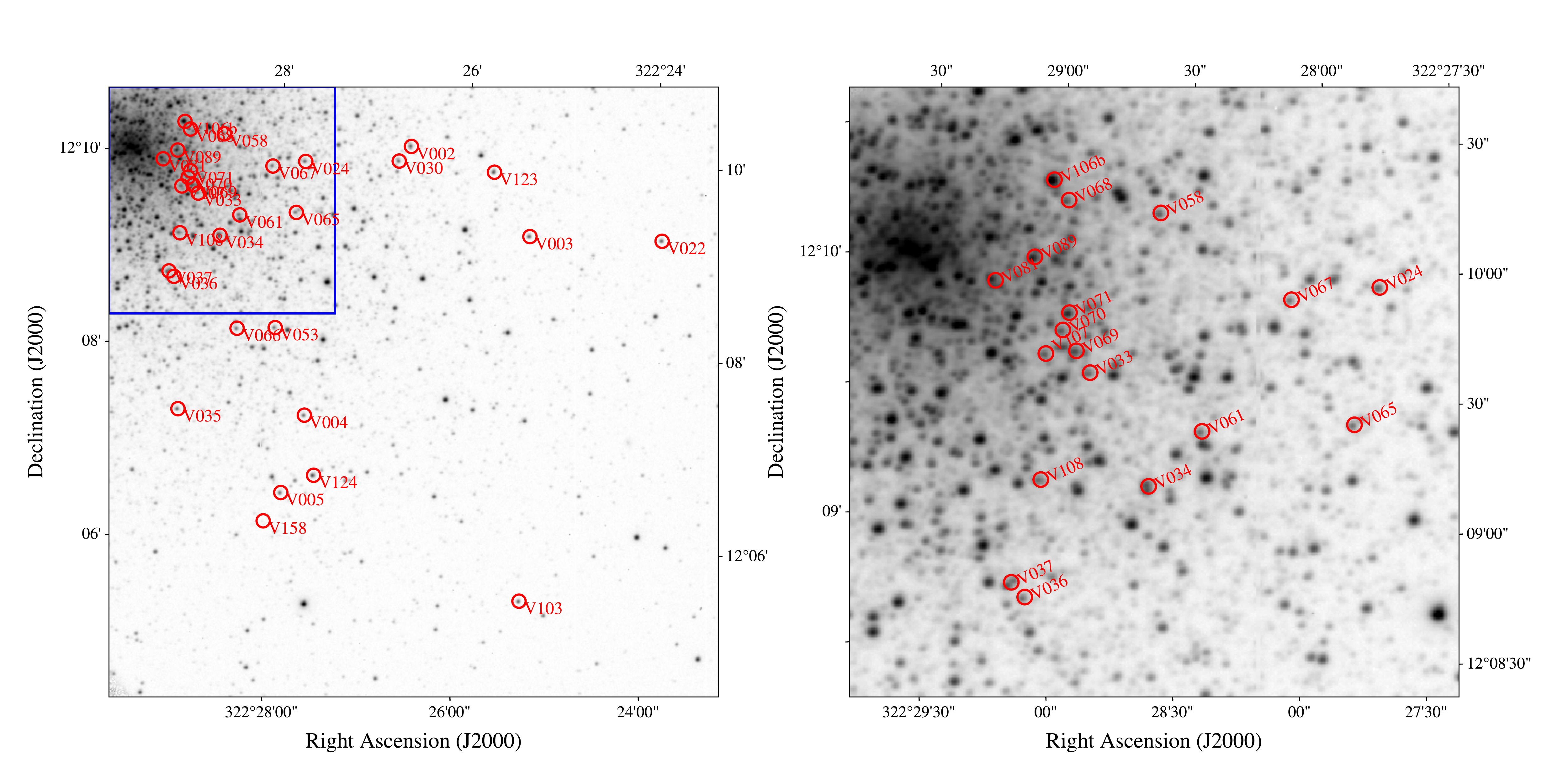}
            \caption{Field M15\_2. Conventions are the same as in Figure~\ref{fig: M15_1_coord}.}
            \label{fig: M15_2_coord}
        \end{figure}    
        
        \begin{figure}
            \centering
            \includegraphics[width=\linewidth]{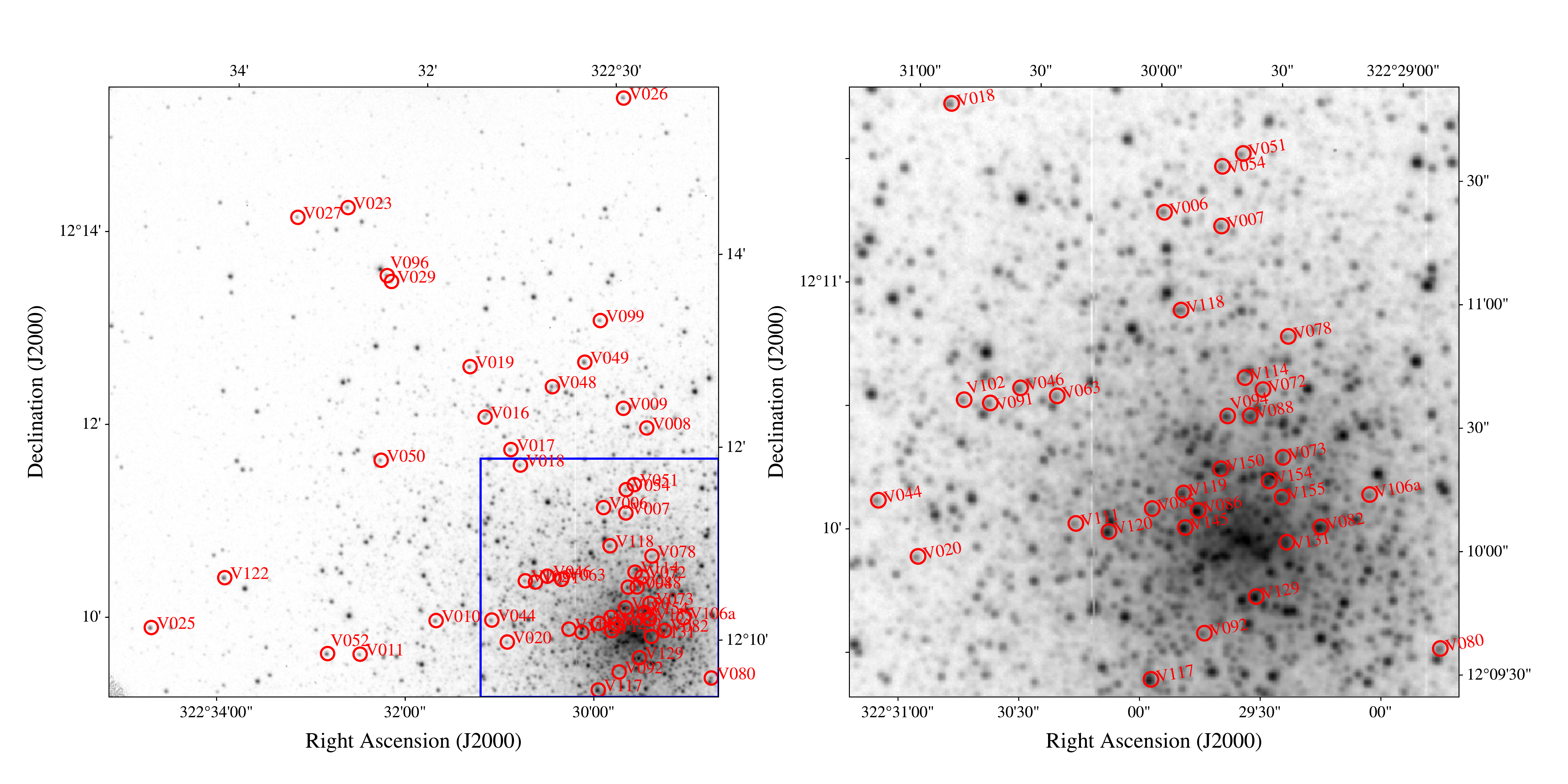}
            \caption{Field M15\_3. Conventions are the same as in Figure~\ref{fig: M15_1_coord}.}
            \label{fig: M15_3_coord}
        \end{figure}
        
        \begin{figure}
            \centering
            \includegraphics[width=\linewidth]{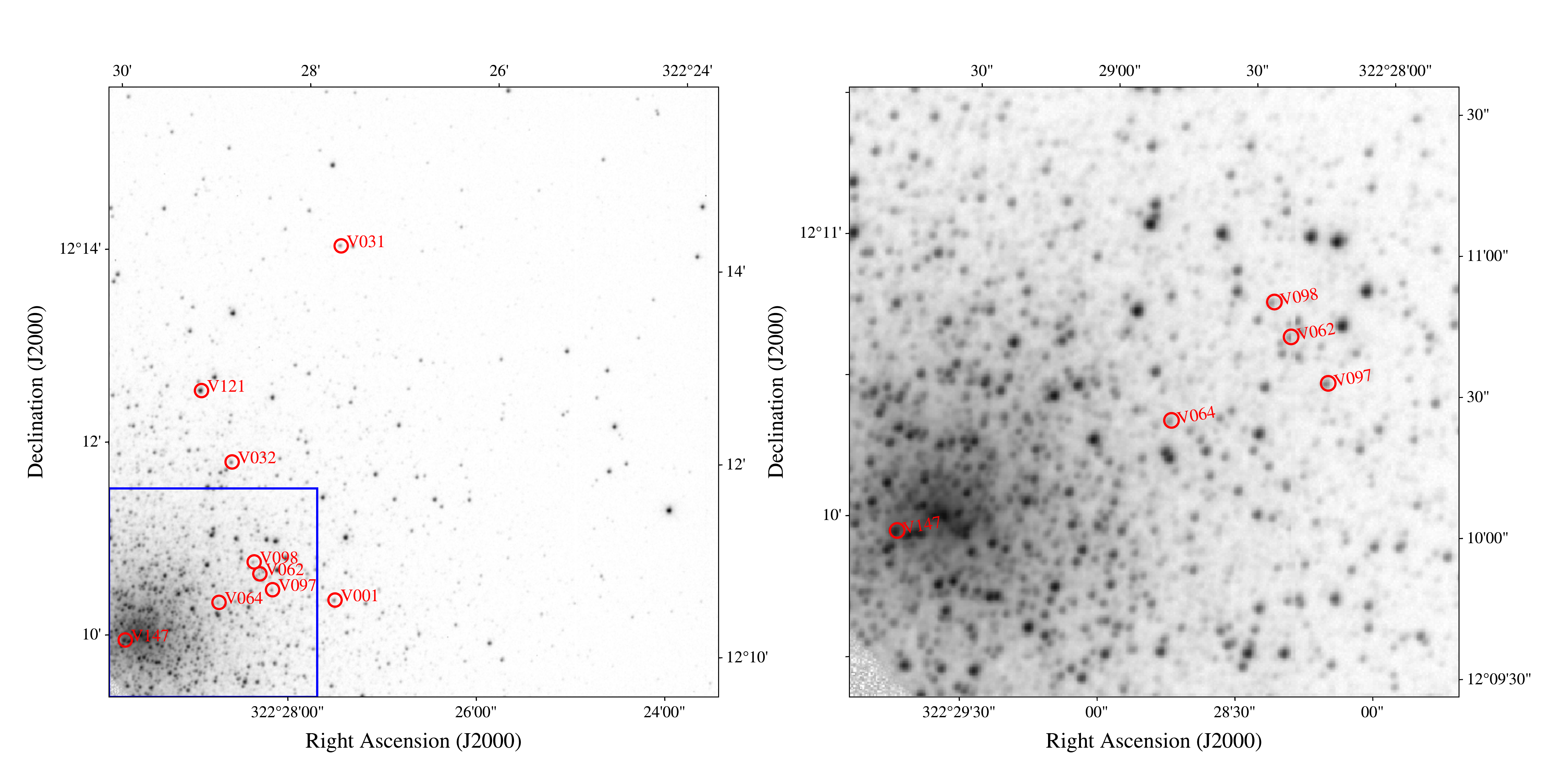}
            \caption{Field M15\_4. Conventions are the same as in Figure~\ref{fig: M15_1_coord}.}
            \label{fig: M15_4_coord}
        \end{figure}


\bsp	
\label{lastpage}
\end{document}